\documentclass[sigconf]{acmart}
\pdfoutput=1

\AtBeginDocument{%
  \providecommand\BibTeX{{%
    \normalfont B\kern-0.5em{\scshape i\kern-0.25em b}\kern-0.8em\TeX}}}

\setcopyright{iw3c2w3}
\copyrightyear{2021}
\acmYear{2021}
\acmDOI{10.1145/3442381.3449788}

\acmConference[WWW '21]{Proceedings of the Web Conference 2021}{April 19--23, 2021}{Ljubljana, Slovenia}
\acmBooktitle{Proceedings of the Web Conference 2021 (WWW '21), April 19--23, 2021,
Ljubljana, Slovenia}
\acmPrice{}
\acmISBN{978-1-4503-8312-7/21/04}

\usepackage[utf8]{inputenc} 
\usepackage[T1]{fontenc}    
\usepackage{amsmath}
\usepackage{hyperref}       
\usepackage{url}            
\usepackage{booktabs}       
\usepackage{amsfonts}       
\usepackage{nicefrac}       
\usepackage{microtype}      
\usepackage{graphicx}
\usepackage{enumitem}
\usepackage{multirow}
\usepackage{subfigure}
\usepackage{bm}

\begin{document}


\title{Disentangling User Interest and Conformity for Recommendation with Causal Embedding}

\author{Yu Zheng$^{1,2}$, Chen Gao$^{1,2}$, Xiang Li$^{3}$, Xiangnan He$^{4}$, Depeng Jin$^{1,2}$, Yong Li$^{1,2\dagger}$}
\affiliation{%
 \institution{$^1$Beijing National Research Center for Information Science and Technology}
 \institution{$^2$Department of Electronic Engineering, Tsinghua University}
 \institution{$^3$University of Hong Kong}
 \institution{$^4$University of Science and Technology of China}
 \institution{$\dagger$Corresponding Author: liyong07@tsinghua.edu.cn}
}
\renewcommand{\authors}{Yu Zheng, Chen Gao, Xiang Li, Xiangnan He, Depeng Jin, Yong Li}

\renewcommand{\shortauthors}{Zheng, et al.}
\begin{abstract}
  Recommendation models are usually trained on observational interaction data. However, observational interaction data could result from users' conformity towards popular items, which entangles users' real interest. Existing methods tracks this problem as eliminating popularity bias, e.g., by re-weighting training samples or leveraging a small fraction of unbiased data. However, the variety of user conformity is ignored by these approaches, and different causes of an interaction are bundled together as unified representations, hence robustness and interpretability are not guaranteed when underlying causes are changing. In this paper, we present DICE, a general framework that learns representations where interest and conformity are structurally disentangled, and various backbone recommendation models could be smoothly integrated. We assign users and items with separate embeddings for interest and conformity, and make each embedding capture only one cause by training with cause-specific data which is obtained according to the colliding effect of causal inference. Our proposed methodology outperforms state-of-the-art baselines with remarkable improvements on two real-world datasets on top of various backbone models. We further demonstrate that the learned embeddings successfully capture the desired causes, and show that DICE guarantees the robustness and interpretability of recommendation.
\end{abstract}



\maketitle

\section{Introduction}\label{sec::intro}

Recent years have witnessed great success in recommender systems, which provide users with personalized contents by mining user preference from the observational interaction data \cite{DLRM19}. However, observational interaction data exhibits strong popularity bias \cite{canamares2018should}, which entangles users' real interest. A user might click an item simply because many other users have clicked it, e.g. in e-commerce platforms items are often displayed with their sales values. In fact, those interactions are mainly driven by users' \textit{conformity}, rather than real \textit{interest}. As a crucial factor for decision making, conformity describes how users tend to follow other people. Meanwhile, conformity towards an item varies according to different users. In order to capture users' pure interest that is independent with conformity, existing approaches track this problem as eliminating popularity bias, a static and global term from the perspective of items, while ignoring the variety of users' conformity.  For example, a sport lover purchases a bicycle with high sales value due to his unique tastes on specific characteristics (e.g. tire size or speed capacity), while an office worker might purchase the same bicycle only because of its high sales. Using uniform popularity bias fails to distinguish these two users' different conformity, since popularity score of an item will be the same for all users. Therefore, disentangling user interest and conformity is crucial enhance recommendation quality.

In this work, we take a different approach from user's perspective. Instead of eliminating popularity bias from the perspective of \textbf{items}, we propose to decompose the observed interactions into two factors in the \textbf{user} side, interest and conformity, and learn disentangled representations for them. Disentangling these two factors is challenging and has not been well explored. Specifically, we face three key challenges. First, conformity depends on both user and item. One user's conformity varies on different items, as well as conformity towards one item from different users. Thus, a scalar bias term for user or item is insufficient, as adopted by existing algorithms \cite{bell2008bellkor}. Second, learning disentangled representations is intrinsically difficult, especially when only observational interaction data is available. In other words, we only have access to the \textit{effect}, but not the \textit{causes}, since there is no labeled ground-truth value for interest and conformity. Third, a click interaction can come from one or both causes of interest and conformity. Therefore, it requires careful design to aggregate and balance the two causes.

\begin{figure}[t]
    \centering
    \includegraphics[width=\linewidth]{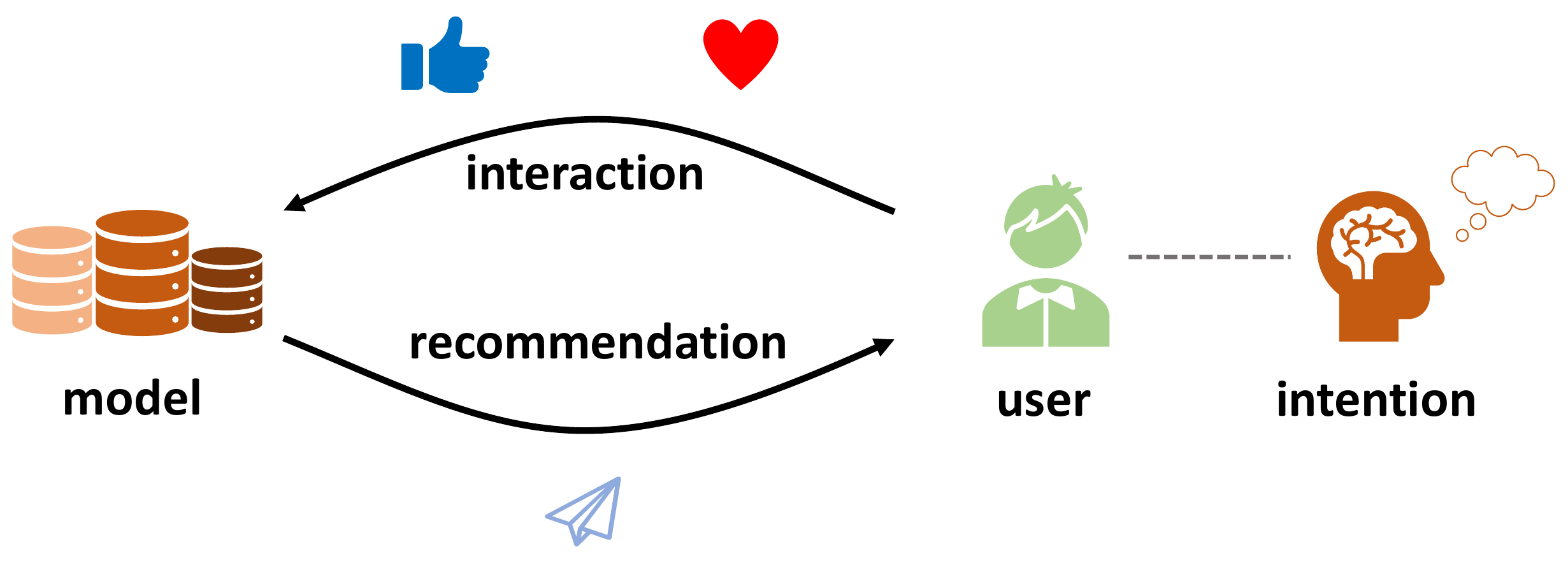}
    \vspace{-20px}
    \caption{Feedback loop of recommender system. Users interact with the model according to user intention, and the model is trained with users' interaction data.}
    \vspace{-20px}
    \label{fig::loop}
\end{figure}

\begin{figure}[t]
    \centering
    \includegraphics[width=\linewidth]{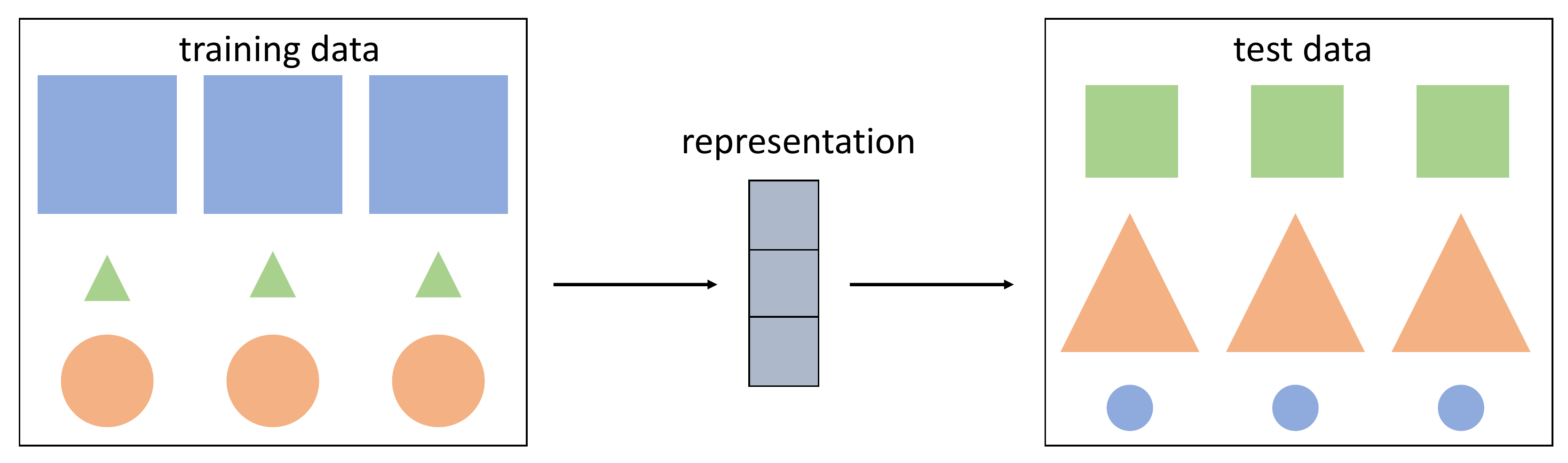}
    \vspace{-20px}
    \caption{Shape recognition under non-IID circumstance. A model can predict shapes from size or color in training data, since rectangles are blue and large, triangles are green and small, circles are orange and medium size. However, under non-IID circumstances as in test data, color and size are different from training data, hence only models that disentangle underlying factors can survive in this example.}
    \vspace{-20px}
    \label{fig::cv}
\end{figure}

Although it is a challenging task, learning disentangled representations of interest and conformity has two main advantages over approaches that only learn a unified embedding for a user or item:

\noindent (1) \textbf{Robustness}. Real-world recommender systems are often trained and updated continuously using real-time user interactions, which forms a feedback loop as illustrated in Figure \ref{fig::loop}, with training data and test data NOT independent and identically distributed (IID) \cite{chaney2018algorithmic}. Causal modeling on the effect (click) and cause (interest and conformity) can lead to more robust models, with stronger generalization ability, especially in non-IID situations where underlying causes are changing \cite{scholkopf2019causality}.

\noindent (2) \textbf{Interpretability}. Interpretable recommendation can benefit both users and platforms of recommender systems, since it improves user-friendliness and facilitates algorithm developing. By disentangling underlying causes, each recommendation score is decomposed as an aggregation of interest score and conformity score. Therefore, explanations towards the two causes can be easily made according to corresponding scores.

In this paper, we present a general framework for \textbf{D}isentangling \textbf{I}nterest and \textbf{C}onformity with \textbf{C}ausal \textbf{E}mbedding (DICE). To capture the variety of conformity, we propose to learn comprehensive embeddings separately for conformity, which is independent with user interest. Instead of using simple scalar popularity value as existing approach, we develop particular methodologies in order to learn disentangled representations for interest and conformity. Specifically, we describe the causal model of how each interaction data is generated. Based on the causal model, we propose particular negative sampling strategies for specific causes based on the \textit{colliding effect} of causal inference \cite{peters2017elements, pearl2018book}, and learn separate embeddings for interest and conformity with cause-specific data. Meanwhile, we add direct supervision on the disentanglement between the two parts of embeddings. To generate final recommendation considering both user interest and conformity, we exploit multi-task and curriculum learning, which successfully balances the two causes.

We evaluate the proposed method on two large-scale benchmark datasets collected from real-world applications. Experimental results show that DICE outperforms state-of-the-art baselines with over 15\% improvements in terms of Recall and NDCG. To investigate the robustness of DICE, we extract test data that is non-IID with training data by conducting \textit{intervention} on conformity. We demonstrate that DICE consistently beats baseline methods under non-IID situations. Furthermore, we provide analytical results on the quality of learned embeddings, which illustrate superior interpretability of the proposed method.

In summary, the main contributions of this paper are as follows:
\vspace{-0.1cm}
\begin{itemize}[leftmargin=*]
	\item To the best of our knowledge, this is the first work to formulate the problem of disentangling user interest and conformity for recommender systems. We tackle the causal recommendation problem from the perspective of users and show that disentangling these two factors is essential for recommender systems, with respect to robustness and interpretability.
	\item We propose a general framework to disentangle interest and conformity. Separate embeddings are adopted to capture the two causes, and different embeddings are trained with cause-specific data, forced to capture only one desired cause. Moreover, we exploit multi-task learning and curriculum learning to balance the two causes.
	\item Extensive experiments are conducted on two large-scale datasets of real-world recommender systems. Results show that DICE achieves significant improvements over state-of-the-art baseline models. Further analysis demonstrates that DICE shows great robustness under non-IID circumstances, and high interpretability of the learned embeddings is guaranteed in DICE as well.
\end{itemize}

The remainder of this paper is as follows. We first introduce the motivation and formulate the problem in Section \ref{sec::problem}. We then elaborate the proposed DICE framework in Section \ref{sec::method}. We conduct experiments in Section \ref{sec::exp}, after that we discuss related works in Section \ref{sec::related}. Finally, we conclude this paper in Section \ref{sec::conclusion}.

\section{Motivation And Problem Overview}\label{sec::problem}

\begin{figure*}
\centering
\includegraphics[width=0.99\textwidth]{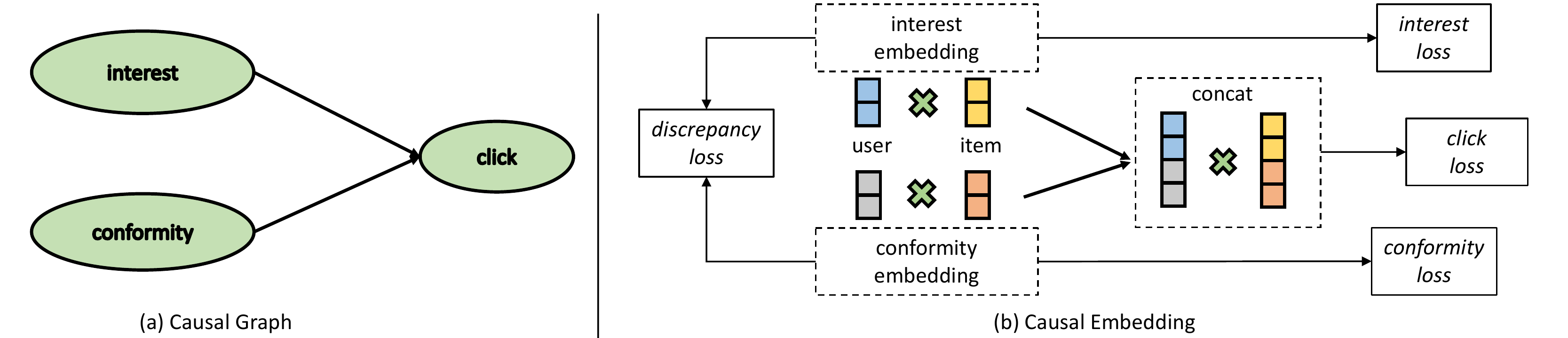}
\vspace{-10px}
\caption{Causal graph and causal embeddings. (a) We make concise causal modeling on each click that it results from two independent causes, interest and conformity. (b) We adopt separate embeddings for interest and conformity, thus each user or item has two embeddings. We force each embedding to capture only one cause by training different embeddings with cause-specific data and adding direct disentanglement supervision, under the framework of multi-task curriculum learning.}
\vspace{-10px}
\label{fig::method}
\end{figure*}

\noindent\textbf{Motivation.}
Algorithms that disentangle underlying semantics have superior generalization ability over \textit{entangled} approaches. Here we focus on a specific form of generalization ability that is not from one data point to another data point in the same distribution, but from one distribution to another distribution. Figure \ref{fig::cv} shows an example of shape recognition, which follows the non-IID condition on training data and test data. Suppose we are developing a shape recognition model, where we learn representations from original pictures and predict their shapes based on the learned representations. It seems like a normal task, however there are traps in it. In fact, models are easily misled by training data, since rectangles are blue and large, triangles are green and small, circles are orange and medium size. As a consequence, a model can predict shapes from color or size, rather than outline. Moreover, if test data is generated from the same distribution (\textit{i.e.} IID with training data), bad models that pay attention to color or size would perform well on test set and we might not even notice what was going wrong. Fortunately, we force training data and test data NOT to be IID as shown in Figure \ref{fig::cv} where color and size are totally different with training data, and evaluate whether our models are robust under this intervened environment. Then only those models that disentangle the underlying semantics (shape, color and size) can survive in our test.

With respect to recommender systems, a click interaction can be triggered by users' real interest, or their conformity towards popular items. In IID situations, it is not necessary for models to distinguish between users' interest and conformity, thus models tend to recommend items according to their popularity values, due to their larger amount of training instances. However, users' conformity at training time and serving time are distinct, since recommender system is a live interactive system shown in Figure \ref{fig::loop}. Therefore, it is essential for recommendation algorithms to be robust in such non-IID situation, especially when underlying causes are different. In this work, we extend conventional causal recommendation algorithms that perform unbiased learning from biased data \cite{joachims2017unbiased}, and propose to disentangle user interest and conformity. Based on recent advancements in causal recommendation \cite{liang2016causal, CausE}, we construct datasets with training data and test data NOT IID. We evaluate the proposed methodology compared with state-of-the-art baseline approaches, and particularly investigate their robustness under non-IID circumstances by interventions.

\noindent\textbf{Problem Formulation.}
Here we formulate the problem of \textit{disentangling user interest and conformity}. Suppose the dataset $\mathcal{O}$ is composed of $N$ instances of $(u, i, p)$, where $p$ is the popularity of item $i$, \textit{i.e.} the number of interactions on item $i$. The distribution of $p$ serves as a proxy for conformity distribution. We first construct intervened test set $\mathcal{O}_{test}$ and normal training set $\mathcal{O}_{train}$, where $D_p^{\mathcal{O}_{test}}$, the distribution of item popularity $p$ in test set, is distinct from that in training set, $D_p^{\mathcal{O}_{train}}$. Our goal is to maximize recommendation performance $\mathcal{R}$, like recall and NDCG, on $\mathcal{O}_{test}$, with models trained on $\mathcal{O}_{train}$ that is NOT IID with $\mathcal{O}_{test}$:

\noindent\textbf{Input}: Observational interaction data $\mathcal{O}$, which is splitted into $\mathcal{O}_{train}$ and $\mathcal{O}_{test}$, with non-IID conditions on popularity distribution $D_p^{\mathcal{O}_{train}}$ and $D_p^{\mathcal{O}_{test}}$.

\noindent\textbf{Output}: A predictive model estimating whether a user will click an item considering both interest and conformity.

\section{DICE: The Proposed Approach}\label{sec::method}

We propose a general framework, named DICE, to learn disentangled representations for interest and conformity. Figure \ref{fig::method} illustrates the holistic design of DICE. To tackle the previously introduced three challenges, our proposed framework is composed of the following three stages: 
\begin{itemize}[leftmargin=*]
    \item \textbf{Causal Embedding}:We propose to utilize separate embeddings instead of scalar values for interest and conformity, to solve the problem of varying conformity.
    \item \textbf{Disentangled Representation Learning}: In order to learn disentangled representations for interest and conformity, we divide training data into cause-specific parts, and train different embeddings with cause-specific data. Direct supervision on embedding distribution is added to reinforce disentanglement.
    \item \textbf{Multi-task Curriculum Learning}: Finally, we develop an easy-to-hard training strategy and exploit curriculum learning to aggregate and balance interest and conformity.
\end{itemize}

\subsection{Causal Embedding}

In this section, we first describe the causal model of how each interaction data is generated from interest and conformity. Then we provide the structural causal model (SCM) and causal graph for click, interest and conformity, based on which we propose to utilize separate embeddings for interest and conformity, solving the first challenge of varying conformity.

\noindent \textbf{How each interaction data is generated?} A click record of a user on an item mainly reflects two aspects: (1) the user's interest in the item’s characteristics, (2) the user's conformity towards the item’s popularity. A click can come from one or both of the two aspects. We propose an additive model to describe how each click record is generated from interest and conformity. Formally, the matching score of a given user $u$ and item $i$ is attained as follow:

\begin{equation}
	S_\mathrm{ui} = S_{ui}^\mathrm{interest} + S_{ui}^\mathrm{conformity},\label{eq::additive}
\end{equation}
where $S_{ui}$ represents the overall matching score, while $S_{ui}^\mathrm{interest}$ and $S_{ui}^\mathrm{conformity}$ stand for a specific cause. This additive model is justified because users tend to have both particularity and conformity when interacting with recommender systems \cite{liu2016you}. Meanwhile, additive models are widely adopted in causal inference and have been shown effective in a bunch of applications \cite{peters2017elements}. In addition, multiplicative model is also adopted in related literature \cite{yang2018unbiased}, which decomposes the click probability as production of exposure probability and the conditional click probability given exposure. However, such multiplicative model entangles interest and conformity from the perspective of user, since users' conformity still takes effect given the exposed items. It is worthwhile to notice that there may be causes other than interest or conformity that lead to a click interaction, but we propose to grasp these two principle factors. Meanwhile, the proposed methodology is a general framework which can be extended to scenarios with multiple causes.

\noindent\textbf{SCM and causal graph for click, interest and conformity}. 
Based on our proposed causal model in (\ref{eq::additive}), we now provide the SCM of our proposed DICE framework, $\zeta_{\mathrm{DICE}}$, along with causal graph in Figure \ref{fig::method}(a):
\begin{equation}
	\begin{aligned}
	&X^{int}_{ui} \coloneq f_1(u, i, N^{int}), \\
	&X^{con}_{ui} \coloneq f_2(u, i, N^{con}), \\
	&Y^{click}_{ui} \coloneq f_3(X^{int}_{ui}, X^{con}_{ui}, N^{click}),
	\end{aligned}
	\label{eq::scm}
\end{equation}
where $N^{int}$, $N^{con}$ and $N^{click}$ are independent noises. SCM $\zeta_{\mathrm{DICE}}$ expresses causal dependence of interest, conformity and click, where $f_1$, $f_2$ and $f_3$ are the underlying causal mechanisms for interest $X^{int}$, conformity $X^{con}$, and click $Y^{click}$ respectively. Practically, those causal mechanisms are decided by optimizing within a given family of functions \cite{peters2017elements}, such as deep neural networks. When we consider interventions on user conformity, we simply replace $X^{con}_{ui}$ with pre-assigned values.

SCM $\zeta_{\mathrm{DICE}}$ in (\ref{eq::scm}) explains the logic on how effect (click) is generated from causes (interest and conformity). However, particular forms of function families in $f_1$, $f_2$ and $f_3$ are still to be determined. As introduced previously, conformity of different users towards the same item are diverse, and so does conformity of the same user towards different items. In other words, conformity depends on both users and items, as well as interest. Therefore, function families for $f_1$ and $f_2$ should better support such flexibility in interest and conformity. We now introduce our proposed design using separate embeddings.

\noindent\textbf{Separate embeddings for interest and conformity} In the proposed DICE framework, we adopt two sets of embeddings to separately capture interest and conformity, instead of using scalar popularity values as in existing approach \cite{bell2008bellkor}, since scalar values are insufficient to capture the diversity of user conformity. As shown in Figure \ref{fig::method}(b), each user has an interest embedding $\bm{u}^{(\mathrm{int})}$ and a conformity embedding $\bm{u}^{(\mathrm{con})}$, and each item also has $\bm{i}^{(\mathrm{int})}$ and $\bm{i}^{(\mathrm{con})}$ for the two causes\footnote{Formally, users should have interest and conformity embeddings, and items should have characteristic and popularity embeddings. We overload the usage of interest and conformity on items to simplify expressions and avoid confusions.}. We use inner product to compute matching score for both causes. Based on the additive causal model in Equation (\ref{eq::additive}), we sum up the two matching scores from corresponding causes, to estimate the overall score on whether a user will click an item. Therefore, the recommendation score for user $u$ and item $i$ is formulated as:
\begin{equation}
\begin{aligned}
        &s^{int}_{ui} = \langle \bm{u}^{(\mathrm{int})}, \bm{i}^{(\mathrm{int})}\rangle, 
        &s^{con}_{ui} = \langle \bm{u}^{(\mathrm{con})}, \bm{i}^{(\mathrm{con})}\rangle, \\
    	&s^{click}_{ui} = s^{int}_{ui} + s^{con}_{ui},
\end{aligned}
\label{eq::overall}
\end{equation}
where $\langle \cdot,\cdot\rangle$ means inner product of two embeddings. Figure \ref{fig::method}(b) demonstrates the disentanglement design of interest embeddings and conformity embeddings. From the perspective of SCM, we restrict the family of functions for $f_1$ and $f_2$ as inner product between two sets of learnable embeddings, and set $f_3$ as a concise additive model which is commonly used in practice \cite{peters2017elements}. By optimizing in two high-dimensional spaces rather than finding the optimal scalar value in 1-D space like existing solutions, diversity of user conformity can be automatically captured in the proposed DICE framework, hence we solve the first challenge.

\subsection{Disentangled Representation Learning}

In this section, we elaborate our designs on disentangling the two causal embeddings for interest and conformity. We propose to train different embeddings with cause-specific data, and decompose the problem into four tasks of conformity modeling, interest modeling, estimating clicks and an extra discrepancy task.

\noindent\textbf{Mining Cause-specific Data}. Disentanglement between interest embedding and conformity embedding means that each embedding captures only one factor, and squeezes out the second factor to the other embedding. To achieve such target, a common and reasonable idea is to train different embeddings with cause-specific data. However, we only have access to the \textit{effect}, which is the observational click data, but we are nearly ignorant on whether the click is caused by interest or conformity. In other words, an equality in (\ref{eq::additive}) is insufficient to recover interest and conformity, because there are infinite solutions to it when there are no ground-truth for two addends and only the summation is available. Therefore, we explore from observational interactions and discover cause-specific data which means those interactions are sourced from individual cause with high probability. The cause-specific data paves the way for disentangling the two underlying causes, interest and conformity.

We first introduce several notations. We use $M^I$ to denote the matrix of interest matching score for all users and items, and $M^C$ for conformity matching score. $M^I$ and $M^C$ are in $\mathbb{R}^{M \times N}$, where $M$ and $N$ are the number of users and items. 
In the causal graph in Figure \ref{fig::method}(a), the three nodes form a \textit{immorality}, and click is the \textit{collider} of interest and conformity \cite{peters2017elements, pearl2018book}.
In fact, the two causes of a collider are independent variables. 
However, if we condition on the collider, the two causes become correlated with each other, and we call it \textit{colliding effect}.
For example, whether a man is popular depends on both his appearance and temper.
Appearance and temper are usually independent, and popularity is the collider of appearance and temper (appearance $\rightarrow$ popularity $\leftarrow$ temper).
Given a popular man who is not good-looking, then he is good-tempered with high probability.
Similarly, an unpopular but good-looking man is most likely bad-tempered.
Therefore, in our task of disentangling interest and conformity, the colliding effect can be utilized to obtain samples that are mostly resulted from one cause.
Specifically, we emphasize on two particular cases in $M^I$ and $M^C$ which are cause-specific:

\noindent\textit{Case 1: The negative item is less popular than the positive item}. If a user $u$ clicks a popular item $a$, while does not click an unpopular item $b$, then we are not sure whether the user’s interest on $a$ is stronger than $b$, because users have conformity towards popular items. In other words, the click can come from the second cause (conformity). Meanwhile, we can also safely conclude that the overall score of the two causes of $a$ is larger than $b$. Hence we have two inequalities in this case:
\begin{equation}
\begin{aligned}
	&M_{ua}^C > M_{ub}^C, \\
	&M_{ua}^I + M_{ua}^C > M_{ub}^I + M_{ub}^C.
\end{aligned}
\label{eq::pop}
\end{equation}

\noindent\textit{Case 2: The negative item is more popular than the positive item}. However, if a user clicks an unpopular item $c$, while does not click a popular item $d$, then the colliding effect can bring more information. Since $c$ is less popular than $d$ which serves as a reasonable proxy for conformity, the click on $c$ is largely due to the user’s interest. Therefore, we have three inequalities in this case, with one extra inequality on interest than the previous case:
\begin{equation}
\begin{aligned}
	&M_{uc}^I > M_{ud}^I, M_{uc}^C < M_{ud}^C, \\
	&M_{uc}^I + M_{uc}^C > M_{ud}^I + M_{ud}^C.
\end{aligned}
\label{eq::unpop}
\end{equation}

We use $\mathcal{O}$ to denote all training instances, which is divided into $\mathcal{O}_1$ and $\mathcal{O}_2$. Specifically, $\mathcal{O}_1$ denotes those instances where negative samples are less popular than positive samples, and $\mathcal{O}_2$ denotes the opposite cases. Correspondingly, $\mathcal{O}_1$ contains the data that inequalities of (\ref{eq::pop}) in \textit{Case 1} hold true, thus $\mathcal{O}_1$ can be utilized to learn conformity and click. $\mathcal{O}_2$ contains the data that fits \textit{Case 2}, hence it can be exploited to learn interest, conformity and click.

By extending one equality to multiple inequalities, we transform the problem from learning absolute values to learning relative relations, which makes the task of disentangling interest and conformity solvable. Specifically, based on these derived inequalities, we obtain user-item interactions that mainly result from one specific cause, and leverage these interactions to optimize corresponding embeddings. Take the famous matrix factorization algorithm for recommendation as an example, usually we  optimize a user embedding matrix and an item embedding matrix to best regress the original interaction matrix, say $M^{click}$. This classical approach unifies all possible causes into one bundled representation for a user or item, thus different causes are entangled, leading to inferior robustness and interpretability under non-IID circumstances, which is quite common in recommender systems. Moreover, debias algorithms such as IPS can not fully solve this problem, since they still adopt unified representations. In contrast to existing approaches, we first decompose the original click matrix $M^{click}$ into two cause-specific matrices, $M^I$ and $M^C$, for interest and conformity respectively. Then two sets of embeddings are adopted, in order to capture interest and conformity separately, and they are further combined to regress click. Different causes are thus disentangled, which achieves better robustness under interventions. We now introduce our causal learning methodology.

With cause-specific data $\mathcal{O}_1$ and $\mathcal{O}_2$, it is possible to model interest and conformity separately. Meanwhile, we propose to estimate click behaviors by combining the two causes, which is the main task for recommendation. Moreover, we further add a discrepancy task in order to make the two sets of embeddings independent with each other, which enhances disentanglement. Therefore, we decompose the problem of disentangling interest and conformity into four tasks, which are conformity modeling, interest modeling, estimating clicks and discrepancy task. We utilize BPR \cite{rendle2012bpr} to model the pairwise quantitative relations in (\ref{eq::pop}) and (\ref{eq::unpop}). Each positive sample is paired with certain number of negative samples, and each training instance is a triplet $(u,i,j)$ containing user ID, positive item ID and negative item ID. We now introduce the four tasks in sequence.

\noindent\textbf{Conformity Modeling} For instances in both $\mathcal{O}_1$ and $\mathcal{O}_2$, we have inequalities for conformity modeling, which are the inequalities for $M^C$. Notice that the directions of inequality are different in the two cases. We use these conformity-specific data to optimize conformity embeddings. BPR loss function is exploited to regress $M^C$ with conformity embeddings. Therefore, the loss function for conformity modeling is formulated as:
\begin{equation}
\begin{aligned}
	L_\mathrm{conformity}^{\mathcal{O}_1} &= \sum_{(u,i,j) \in \mathcal{O}_1}\mathrm{BPR}(\langle \bm{u}^{(\mathrm{con})}, \bm{i}^{(\mathrm{con})}\rangle, \langle \bm{u}^{(\mathrm{con})}, \bm{j}^{(\mathrm{con})}\rangle), \\
	L_\mathrm{conformity}^{\mathcal{O}_2} &= \sum_{(u,i,j) \in \mathcal{O}_2}-\mathrm{BPR}(\langle \bm{u}^{(\mathrm{con})}, \bm{i}^{(\mathrm{con})}\rangle, \langle \bm{u}^{(\mathrm{con})}, \bm{j}^{(\mathrm{con})} \rangle), \\
	L_\mathrm{conformity}^{\mathcal{O}_1+\mathcal{O}_2} &= L_\mathrm{conformity}^{\mathcal{O}_1} + L_\mathrm{conformity}^{\mathcal{O}_2}.
\end{aligned}
\end{equation}

\noindent\textbf{Interest Modeling} In $\mathcal{O}_2$, negative items are more popular than positive items, and those interactions are largely due to users' interest. These data is interest-specific, and we have inequalities for interest modeling. We also use BPR to optimize interest embeddings to learn such pairwise preference, in order to regress $M^I$. The loss function only takes effect for instances in $\mathcal{O}_2$:
\begin{equation}
	L_\mathrm{interest}^{\mathcal{O}_2} = \sum_{(u,i,j) \in \mathcal{O}_2}\mathrm{BPR}(\langle \bm{u}^{(\mathrm{int})}, \bm{i}^{(\mathrm{int})}\rangle, \langle \bm{u}^{(\mathrm{int})}, \bm{j}^{(\mathrm{int})}\rangle).
\end{equation}

\noindent\textbf{Estimating Clicks} This is the main target for recommender systems, and we combine the two causes to estimate clicks as introduced in (\ref{eq::overall}), with a concise additive model. For each instance in training set $\mathcal{O}$, which is the union of $\mathcal{O}_1$ and $\mathcal{O}_2$, we use BPR to maximize the margin between scores of positive items and negative items, so as to regress $M^{click}$. The loss function for click estimation is thus formulated as follow:
\begin{equation}
	L_\mathrm{click}^{\mathcal{O}_1+\mathcal{O}_2} = \sum_{(u,i,j) \in \mathcal{O}}\mathrm{BPR}(\langle \bm{u}^t, \bm{i}^t\rangle, \langle \bm{u}^t, \bm{j}^t\rangle).
\end{equation}
$\bm{u}^t$, $\bm{i}^t$ and $\bm{j}^t$ are concatenation of interest embedding and conformity embedding for user and item:
\begin{equation}
	\bm{u}^t = \bm{u}^{(\mathrm{int})}\|\bm{u}^{(\mathrm{con})}, \bm{i}^t = \bm{i}^{(\mathrm{int})}\|\bm{i}^{(\mathrm{con})}, \bm{j}^t = \bm{j}^{(\mathrm{int})}\|\bm{j}^{(\mathrm{con})},
\end{equation}
where $\|$ means concatenation of two embeddings. We use the concatenation form here for simplicity, which is equivalent to the summation form in (\ref{eq::overall}). The BPR loss pushes the recommendation score for the positive item $i$ to be higher than the negative item $j$.

Interest modeling and conformity modeling disentangle the two causes by training different embeddings with different cause-specific data. Meanwhile, the main task on estimating clicks also strengthens this disentanglement as a constraint. For example, in terms of a training instance $(u, i, j)$ where negative item $j$ is more popular than positive item $i$, interest modeling task forces the two sets of embeddings to learn that user $u$'s interest in $i$ is larger than $j$, and conformity modeling task forces them to learn that user $u$'s conformity towards item $i$' is less than $j$. Meanwhile, estimating clicks forces them to learn that the overall strength on $i$ is larger than $j$. Therefore, what the model really learns is that the advantage of $i$ over $j$ with respect to interest dominates the disadvantage in conformity, which can be best learned by capturing only one cause with one embedding.

\noindent\textbf{Discrepancy Task} Besides the three tasks above that disentangle interest and conformity by optimizing different embeddings with cause-specific data, we impose direct supervision on the embedding distribution to reinforce this disentanglement. Suppose $\mathbf{E}^{(\mathrm{int})}$ and $\mathbf{E}^{(\mathrm{con})}$ represent two sets of embeddings of all users and items. We examine three candidate discrepancy loss functions, which are L1-inv, L2-inv and distance correlation (\textit{dCor}). L1-inv and L2-inv \textit{maximize} L1 and L2 distances between $\mathbf{E}^{(\mathrm{int})}$ and $\mathbf{E}^{(\mathrm{con})}$ respectively. 
We refer to \cite{szekely2009brownian,szekely2007measuring} for details on \textit{dCor}. From high level, \textit{dCor} is a more reasonable choice, since it focuses on the correlations of pairwise distances between interest embeddings and conformity embeddings. 
The three options for discrepancy loss function are $-L1(\mathbf{E}^{(\mathrm{int})}, \mathbf{E}^{(\mathrm{con})})$, $-L2(\mathbf{E}^{(\mathrm{int})}, \mathbf{E}^{(\mathrm{con})})$ and $dCor(\mathbf{E}^{(\mathrm{int})}, \mathbf{E}^{(\mathrm{con})})$. We will compare them in experiments.

Figure \ref{fig::method}(b) illustrates the four decomposed tasks using disentangled embeddings for interest and conformity. By training different embeddings with cause-specific data and imposing direct supervision on the embedding distribution, we solve the second challenge of learning disentangled representations.

\subsection{Multi-task Curriculum Learning}

In the proposed framework, we overcome the last challenge of aggregating interest and conformity by multi-task curriculum learning. To be specific, we train causal embeddings with the four above tasks simultaneously, and combine these loss functions together: 
\begin{equation}
	L = L_\mathrm{click}^{\mathcal{O}_1+\mathcal{O}_2} + \alpha  (L_\mathrm{interest}^{\mathcal{O}_2} + L_\mathrm{conformity}^{\mathcal{O}_1+\mathcal{O}_2}) + \beta L_\mathrm{discrepancy}.
\end{equation}
Since estimating clicks is the main task for recommendation, $\alpha$ and $\beta$ should be less than 1 from intuition. Meanwhile, discrepancy task directly influences the distribution of embeddings, thus too large $\beta$ would negatively impact interest and conformity modeling.

As introduced previously, we obtain two or three inequalities when the negative sample is less or more popular than the positive sample, respectively. Notice that those inequalities will hold true with high probability when the popularity gap is sufficiently large. Therefore, we develop \textbf{P}opularity based \textbf{N}egative \textbf{S}ampling with \textbf{M}argin (PNSM) to guarantee those quantitative relations. Specifically, if the popularity of the positive sample is $p$, then we will sample negative instances from items with popularity larger than $p+m_{up}$, or lower than $p-m_{down}$, where $m_{up}$ and $m_{down}$ are positive margin values. By sampling negative items with popularity margin, we gain high confidence in our causal models. Later experiments show that popularity based negative sampling with margin is of crucial importance for learning disentangled and robust representations.

Inspired by curriculum learning \cite{bengio2009curriculum}, we adopt an easy-to-hard strategy on training DICE by adding decay on margin values and loss weights. Specifically, when margin values $m_{up}$ and $m_{down}$ are large, we have high confidence on those inequalities for interest and conformity modeling, which means the tasks are \textit{easier} and we set high loss weights $\alpha$ for $L_{\mathrm{interest}}$ and $L_{\mathrm{conformity}}$. As we train the model, we increase the difficulty by decaying margin values, as well as loss weights $\alpha$, by a factor of 0.9 after each epoch. With curriculum learning, the proposed approach learns stronger disentanglement for high-confidence samples. Furthermore, this adaptive design also makes the proposed method not sensitive to initial values of hyper-parameters. We will compare the performance of curriculum learning with normal learning in experiments. Interest and conformity are elegantly aggregated by multi-task curriculum learning, hence the last challenge is addressed.

In summary, we propose an additive causal model on user interest and conformity. Based on SCM $\zeta_{\mathrm{DICE}}$, we develop separate causal embeddings for individual cause, which capture the diversity of conformity and interest. A bunch of inequalities are derived from our causal model, decomposing the causal learning task into conformity modeling, interest modeling, estimating clicks and discrepancy task. Disentangled representations for underlying causes are obtained by training different embeddings with cause-specific data. To attain robust recommendation, multi-task curriculum learning is adopted to aggregate the two causes. Meanwhile, our causal framework is based on how data is generated and hence they are model-independent. Therefore, the proposed DICE methodology provides a highly general framework for disentangling user interest and conformity, which can be smoothly integrated into existing recommendation models. In our experiments, we successfully develop DICE on top of state-of-the-art recommender systems based on Graph Convolutional Networks.

\section{Experiments}\label{sec::exp}

\begin{table}[t]
\caption{Statistics of datasets. (Ent. stands for entropy value of the number of interactions for all items. Larger entropy value of test data shows the non-IID condition.}
\vspace{-10px}
\centering
{\small
\begin{tabular}{lccccc}
\toprule
Dataset & User &  Item & Interaction & Ent. Train &  Ent. Test  \\
\midrule
Movielens-10M & 37962 & 4819 & 1371473 & 6.22 & 7.97 \\
Netflix & 32450 & 8432 & 2212690 & 6.85 & 8.54 \\
\bottomrule
\end{tabular}\label{tab::dataset}
}
\vspace{-15px}
\end{table}

\begin{table*}
    \caption{Overall performance on Movielens-10M dataset and Netflix dataset.}
    \vspace{-10px}
    \label{tab::overall}
    \begin{tabular}{c|c|ccc|ccc|ccc|ccc}
      \toprule
      \multicolumn{2}{c|}{\multirow {2}{*}{Dataset}} & \multicolumn{6}{c|}{Movielens-10M} & \multicolumn{6}{c}{Netflix} \\
      \multicolumn{2}{c|}{} & \multicolumn{3}{c}{TopK = 20} & \multicolumn{3}{c|}{TopK = 50} & \multicolumn{3}{c}{TopK = 20} & \multicolumn{3}{c}{TopK = 50} \\
      \hline
      Model & Method & Recall & HR & NDCG & Recall & HR & NDCG & Recall & HR & NDCG & Recall & HR & NDCG \\
      \midrule
      \multicolumn{1}{c|}{\multirow {7}{*}{MF}} & None & 0.1286	& 0.4429	& 0.0846 & 0.2346 & 0.6295 & 0.1170 & \underline{0.1122} & \underline{0.5194} & 0.0943 & 0.1928	& \underline{0.6749} & 0.1185 \\
      \multicolumn{1}{c|}{} & IPS & 0.1335 & 0.4434 & 0.0852 & 0.2376 & 0.6288 & 0.1174 & 0.1058 & 0.4882 & 0.0864 & 0.1855 & 0.6562 & 0.1112 \\
      \multicolumn{1}{c|}{} & IPS-C &  0.1367 & 0.4564 & 0.0875 & 0.2429	 & 0.6383 & 0.1203 & 0.1119 & 0.5046 & 0.0919 & \underline{0.1938} & 0.6700 & 0.1174 \\
      \multicolumn{1}{c|}{} & IPS-CN & \underline{0.1412} & \underline{0.4700} & \underline{0.0925} & \underline{0.2509} & \underline{0.6477} & \underline{0.1264} & 0.1080 & 0.5042 & 0.0935 & 0.1912 & 0.6621 & 0.1185 \\
      \multicolumn{1}{c|}{} & IPS-CNSR & 0.1365	& 0.4588 & 0.0895 & 0.2419 & 0.6366 & 0.1219 & 0.1110 & 0.5159 & \underline{0.0948} & 0.1937 & 0.6713 & \underline{0.1192} \\
      \multicolumn{1}{c|}{} & CausE & 0.1157 & 0.4066 & 0.0744 & 0.2121 & 0.5924 & 0.1037 & 0.0935 & 0.4641 & 0.0782 & 0.1651 & 0.6272 & 0.0994  \\
      \multicolumn{1}{c|}{} & DICE & \bf{0.1634} & \bf{0.5197} & \bf{0.1084} & \bf{0.2872} & \bf{0.6975} & \bf{0.1468} & \bf{0.1258} & \bf{0.5545} & \bf{0.1070} & \bf{0.2164} & \bf{0.7090} & \bf{0.1345} \\
      \hline
      \multicolumn{1}{c|}{\multirow {7}{*}{GCN}} & None & 0.1378	 & 0.4625 & 0.0898 & 0.2513 & 0.6505 & 0.1247 & 0.1026 & 0.4908 & 0.0870 & 0.1842 & 0.6609 & 0.1112 \\
      \multicolumn{1}{c|}{} & IPS & 0.1394 & 0.4645 & 0.0919 & 0.2538 & 0.6473 & 0.1275 & 0.1101 & 0.5091 & 0.0950 & 0.1941 & 0.6657 & 0.1203 \\
      \multicolumn{1}{c|}{} & IPS-C & \underline{0.1478} & \underline{0.4829} & \underline{0.0971} & \underline{0.2654} & \underline{0.6632} & \underline{0.1339} & \underline{0.1157} & \underline{0.5219} & \underline{0.1004} & \underline{0.2037} & \underline{0.6816} & \underline{0.1270} \\
      \multicolumn{1}{c|}{} & IPS-CN & 0.1119 & 0.3997 & 0.0701 & 0.2281 & 0.6112 & 0.1057 & 0.0726 & 0.3991 & 0.0643 & 0.1472 & 0.5841 & 0.0866 \\
      \multicolumn{1}{c|}{} & IPS-CNSR & 0.1300 & 0.4427 & 0.0852 & 0.2336 & 0.6282 & 0.1171 & 0.0826 & 0.4337 & 0.0715 & 0.1589 & 0.6124 & 0.0940 \\
      \multicolumn{1}{c|}{} & CausE & 0.1027 & 0.3729 & 0.0632 & 0.2044	& 0.5811 & 0.0941 & 0.0838 & 0.4289 & 0.0677 & 0.1569 & 0.6119 & 0.0902 \\
      \multicolumn{1}{c|}{} & DICE & \bf{0.1812} & \bf{0.5563} & \bf{0.1228} & \bf{0.3100} & \bf{0.7216} & \bf{0.1629} & \bf{0.1420} & \bf{0.5910} & \bf{0.1217} & \bf{0.2367} & \bf{0.7340} & \bf{0.1499} \\
      \bottomrule
    \end{tabular}
    \vspace{-10px}
\end{table*}

In this section, we conduct experiments to show the effectiveness of the proposed framework. Specifically, we aim to answer the following research questions:

\begin{itemize}[leftmargin=*]
    \item \textbf{RQ1}: How does our proposed DICE framework perform compared with state-of-the-art causal recommendation methods under non-IID circumstances? Particularly, is it necessary to replace scalar bias term with embedding?
    \item \textbf{RQ2}: Can the proposed DICE framework guarantee interpretability and robustness?
    \item \textbf{RQ3}: What is the role of each component in the proposed methodology, including negative sampling, conformity modeling, curriculum learning, and discrepancy loss?
    \item \textbf{RQ4}: What is the effect of intervened data inserted into training set? How does DICE perform when no intervened training data is available?
\end{itemize}

\subsection{Experimental Settings}

\textbf{Datasets} We conduct experiments on two million-scale datasets collected from real-world applications, Movielens-10M dataset \cite{movielens} and Netflix Prize dataset \cite{netflix}, and Table \ref{tab::dataset} lists the statistics of two datasets.

\noindent\textbf{Data Preprocessing} In order to measure the performance of causal learning under non-IID circumstances, intervened test sets are needed, and thus all datasets are transformed following the standard protocol introduced in related literatures \cite{CausE,liang2016causal}. We binarize the datasets by keeping ratings of five stars as one, and others as zero. To conduct intervention on conformity, we randomly sample 40\% of the records with equal probability in terms of items, and leave the other 60\% as training data. In other words, items are sampled with probability as \textbf{inverse} popularity, which means popular items are less selected. Moreover, we cap the probability at 0.9 to limit the number of items that do not show in training set \cite{CausE}. Finally, we obtain a 70/10/20 split for training set (60\% normal and 20\% intervened), validation set (10\% intervened) and test set (20\% intervened). Test data can be regarded as recommendation result under a fully random policy. As a consequence, conformity in test data is distinct from that in training data, since users have access to all items with equal probability in test data, rather than seeing more popular items in training data. We refer to \cite{CausE,liang2016causal} for details on extracting an intervened test set from original interaction data. To show that training data and test data are non-IID, we count the number of interactions for each item and calculate the entropy, hence larger entropy value indicates that different items are of more equal probability to be exposed to users. As illustrated in Table \ref{tab::dataset}, entropy on test data is much larger than that on training data for both datasets. In other words, models are trained on normal data, while evaluated on intervened data.

\noindent\textbf{Recommendation Models} Causal approaches usually serve as additional methods upon backbone recommendation models. We use the most adopted recommendation model, Matrix Factorization (MF) \cite{koren2009matrix} to compare different approaches. Meanwhile, we also incorporate the state-of-the-art collaborate filtering model, Graph Convolutional Networks (GCN) \cite{hamilton2017inductive,ying2018graph,he2020lightgcn}, to investigate whether algorithms generalize across different recommendation models. Specifically, we use BPR-MF \cite{rendle2012bpr} and LightGCN \cite{he2020lightgcn}, which are both state-of-the-art recommendation models. 

\noindent\textbf{Experiment Setups} For IPS based models, we fix the embedding size as 128. While for CausE and DICE, the embedding size is fixed as 64, since they contain two sets of embeddings. Therefore, the number of parameters are the same for all methods to guarantee fair comparison. We set $\alpha$ as 0.1 and $\beta$ as 0.01, which shows great performance and agnostic to both datasets and backbone models in experiments. We use BPR \cite{rendle2012bpr} as the loss function for all baselines. We use Adam \cite{kingma2014adam} for optimization. Other hyper-parameters for our method and baselines are tuned by grid search. The code and data are available at \url{https://github.com/tsinghua-fib-lab/DICE}.

\subsection{Performance Comparison (RQ1)}

\subsubsection{\textbf{Overall Performance}} We compare our approach with the following state-of-the-art causal recommendation methods:
\begin{itemize}[leftmargin=*]
    \item \textbf{IPS} \cite{schnabel2016recommendations,joachims2017unbiased}: IPS eliminates popularity bias by re-weighting each instance according to item popularity. Specifically, weight for an instance is set as the inverse of corresponding item popularity value, hence popular items are imposed lower weights, while the importance for long-tail items are boosted.
    \item \textbf{IPS-C} \cite{bottou2013counterfactual}: This method adds max-capping on IPS value to reduce the variance of IPS.
    \item \textbf{IPS-CN} \cite{gruson2019offline}: This method further adds normalization which also achieved lower variance than plain IPS, at the expense of introducing a small amount of bias.
    \item \textbf{IPS-CNSR} \cite{gruson2019offline}: Smoothing and re-normalization are added to attain more stable output of IPS.
    \item \textbf{CausE} \cite{CausE}: This method requires a large biased dataset and a small unbiased dataset. Each user or item has two embeddings to perform matrix factorization (MF) on the two datasets respectively, and L1 or L2 regularization is exploited to force the two sets of embeddings similar with each other. 
\end{itemize}
We also include simple MF and GCN without using any causal methods for comparison. We evaluate top-k recommendation performance for implicit feedback \cite{rendle2012bpr}, which is the most common setting for recommendation. We adopt three frequently used metrics, which are Recall, Hit Ratio and NDCG.

Results on two datasets are listed in Table \ref{tab::overall}. We have the following observations:

\begin{itemize}[leftmargin=*]
    \item \textbf{Our proposed DICE framework outperforms baselines with significant improvements with respect to all metrics on both datasets.} For example, DICE makes over 15\% improvements with respect to NDCG@50 using MF as backbone on Moveilens-10M dataset, and over 20\% improvements with respect to Recall@20 using GCN as backbone on Netflix dataset. Results show that the disentanglement design of interest embeddings and conformity embeddings successfully distinguish the two causes of user interactions. It allows the framework to capture invariant interest from training data, and adapt to intervened conformity in test cases.
    \item \textbf{DICE is highly general framework which can be combined with various recommendation models.} Besides attaining the best performance on both datasets, the proposed DICE framework also outperforms all other baselines with both recommendation models, MF and GCN. The proposed concise causal model is sourced from how the data is generated, thus the proposed framework is independent with backbone recommendation models. Results based on MF and GCN illustrate that DICE is a general framework, which can be smoothly integrated into various embedding based recommendation algorithms.
    \item \textbf{Entangled causal models are not stable on different datasets and metrics.} From the results in Table \ref{tab::overall}, entangled causal models like IPS and CausE can not make improvements consistently on different datasets and metrics. For example, IPS-CN achieves the second best performance on Movielens-10M dataset, but fails to make improvements on Netflix dataset with MF as recommendation model. In addition, IPS-CNSR attains decent performance with respect to NDCG on Netflix dataset with MF as recommendation model, but it is even worse than None (no causal model) in terms of another metric, HR. Without disentangling interest and conformity, those causal models are not stable on different datasets and metrics. In contrast, the disentangled DICE framework attains consistent improvements by disentangling the underlying causes.
\end{itemize}

\subsubsection{\textbf{Comparison between Embedding and Scalar}} Using a scalar bias term for each item and user is frequently adopted to capture the influence of popularity \cite{bell2008bellkor}. However, it is insufficient to express the diversity of user conformity. For example, user $a$ has stronger conformity towards item $s$ than user $b$, thus bias term for user $a$ should be higher than user $b$. However, user $a$ would have weaker conformity towards item $t$ than user $b$, which requires bias term for user $a$ to be lower than user $b$. It is common in practice since users tend to have different conformity in their familiar and unfamiliar fields, such as categories of items or genres of movies. The above contradiction demonstrates the limited power of using scalar values to capture user conformity. In our work, we propose to exploit embeddings instead of simple scalars. By raising dimensions of solution space, the diversity of user conformity is guaranteed. For example, the above contradiction can be easily resolved by using 2-D vectors for user conformity and item popularity, rather than 1-D scalars.

\begin{figure}[t]
\begin{minipage}[t]{0.49\linewidth}
\centering
\includegraphics[width=\linewidth]{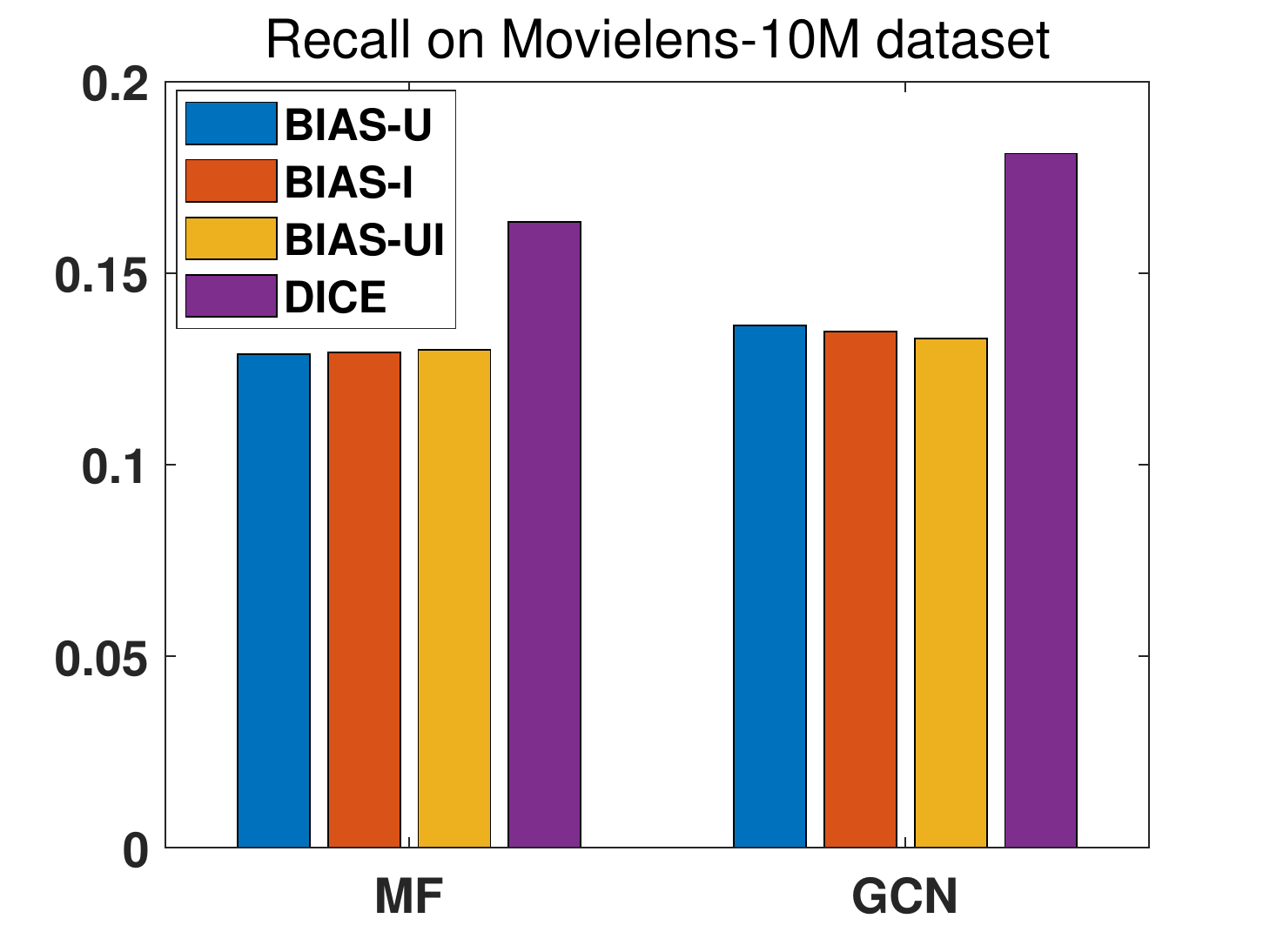}
\end{minipage}
\begin{minipage}[t]{0.49\linewidth}
\centering
\includegraphics[width=\linewidth]{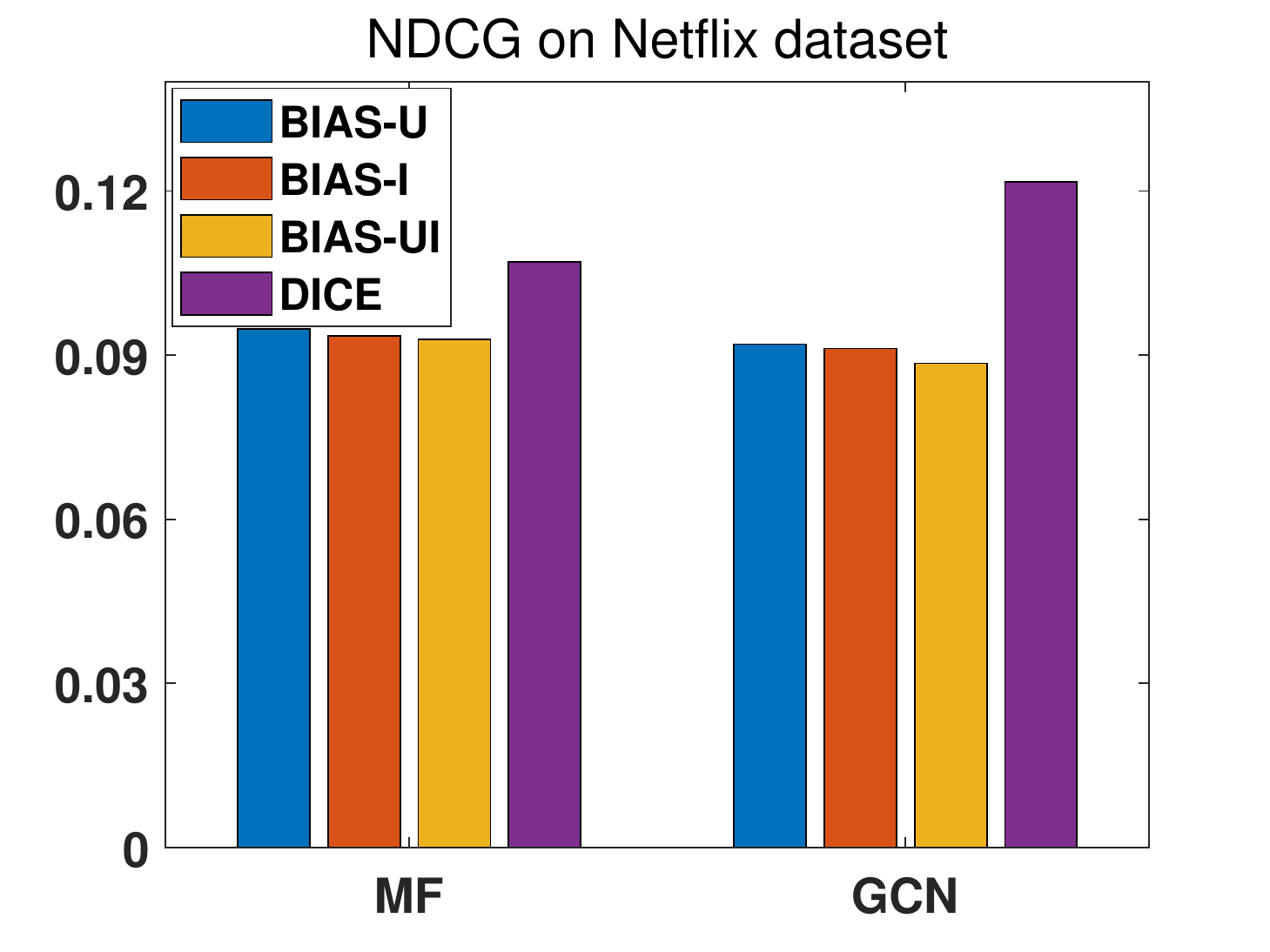}
\end{minipage}
\vspace{-10px}
\caption{Comparison between using embeddings and using scalars on two datasets.}
\vspace{-20px}
\label{fig::bmf}
\end{figure}

We compare the proposed DICE framework using embeddings with existing algorithms using scalar values. We include bias terms for both users and items. Specifically, we compare DICE with BIAS-U (adding scalar bias term for each user), BIAS-I (adding scalar bias term for each item) and BIAS-UI (adding scalar bias term for each user and item) on both MF and GCN. Figure \ref{fig::bmf} shows the results on the two datasets. DICE outperforms all other models with scalar bias terms with significant margin, proving that simple scalar values are insufficient to capture the diversity of user conformity. Experiments on both MF and GCN show that it is necessary to use embeddings rather than scalar values for conformity modeling.

\subsection{Interpretability and Robustness (RQ2)}
As introduced previously, disentangled algorithms are generally more interpretable and robust than entangled competitors. In this section, we investigate whether the proposed DICE framework has such advantages.

\begin{figure*}[t]
\subfigure[]{
\begin{minipage}[t]{0.3\linewidth}
\label{fig:mini:subfig:a}
\includegraphics[width=\linewidth]{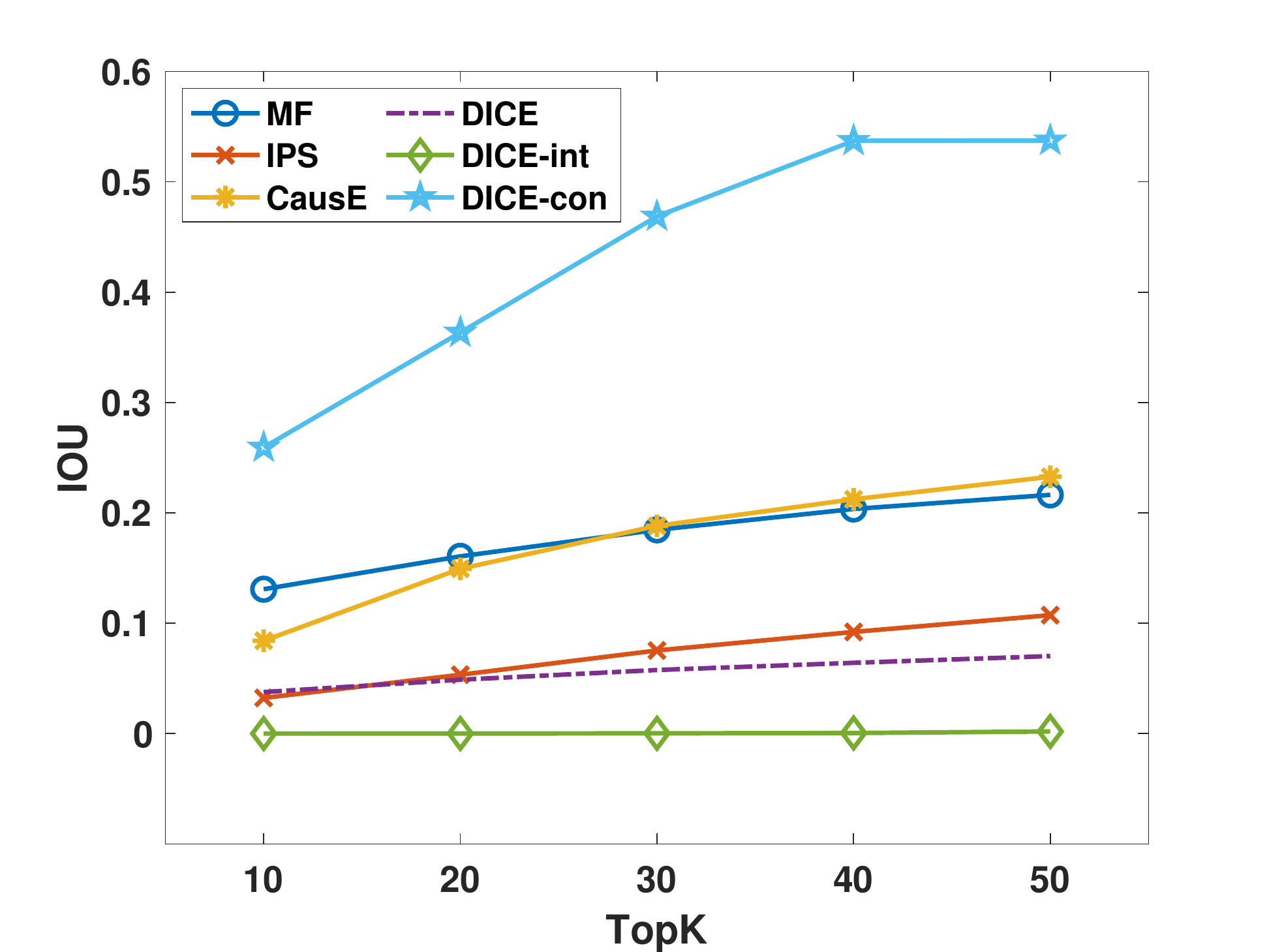}
\label{fig::overlap}
\end{minipage}}
\subfigure[]{
\begin{minipage}[t]{0.6\linewidth}
\label{fig:mini:subfig:b}
\begin{minipage}[t]{0.5\linewidth}
\includegraphics[width=\linewidth]{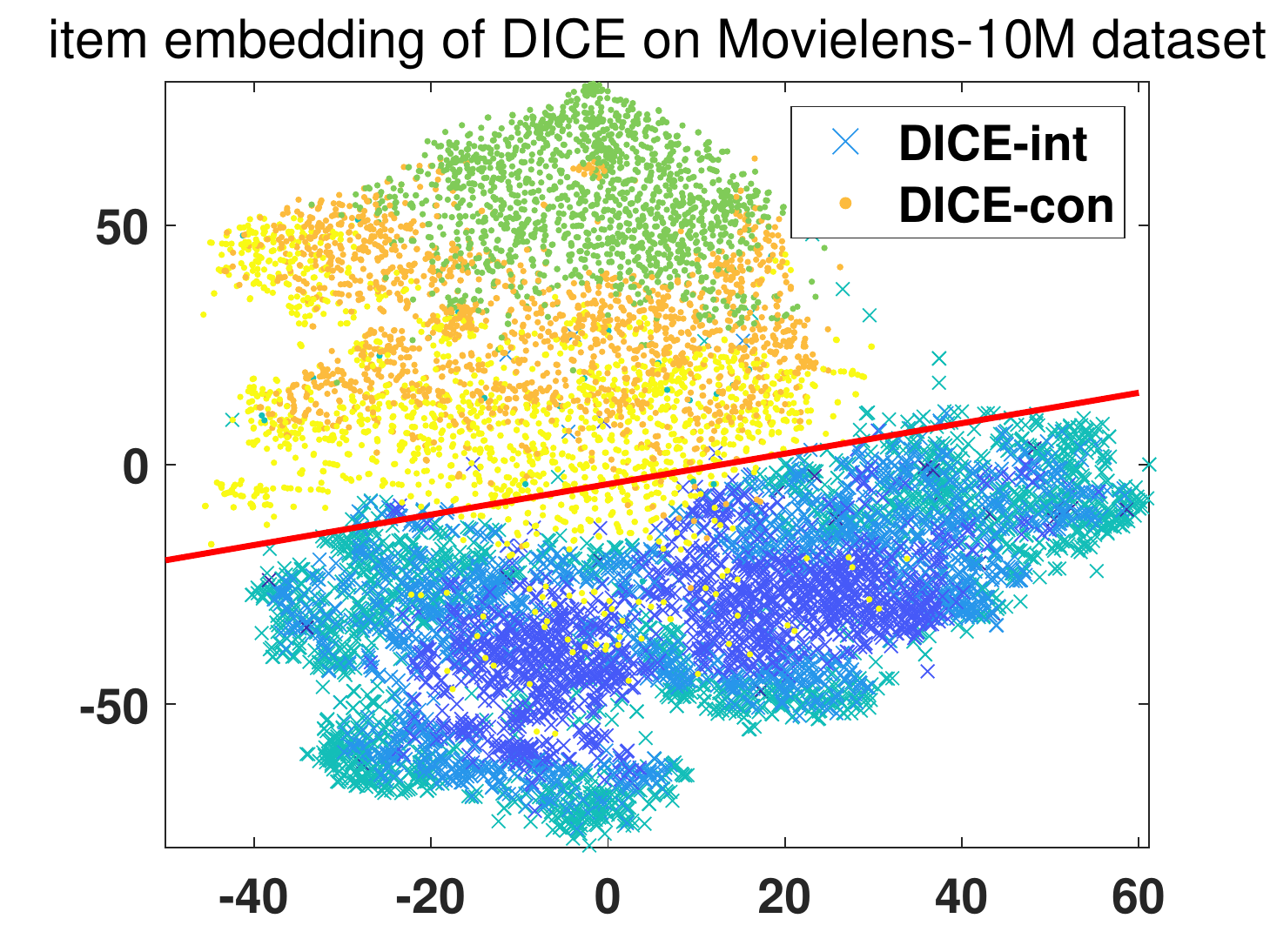}
\end{minipage}
\begin{minipage}[t]{0.5\linewidth}
\includegraphics[width=\linewidth]{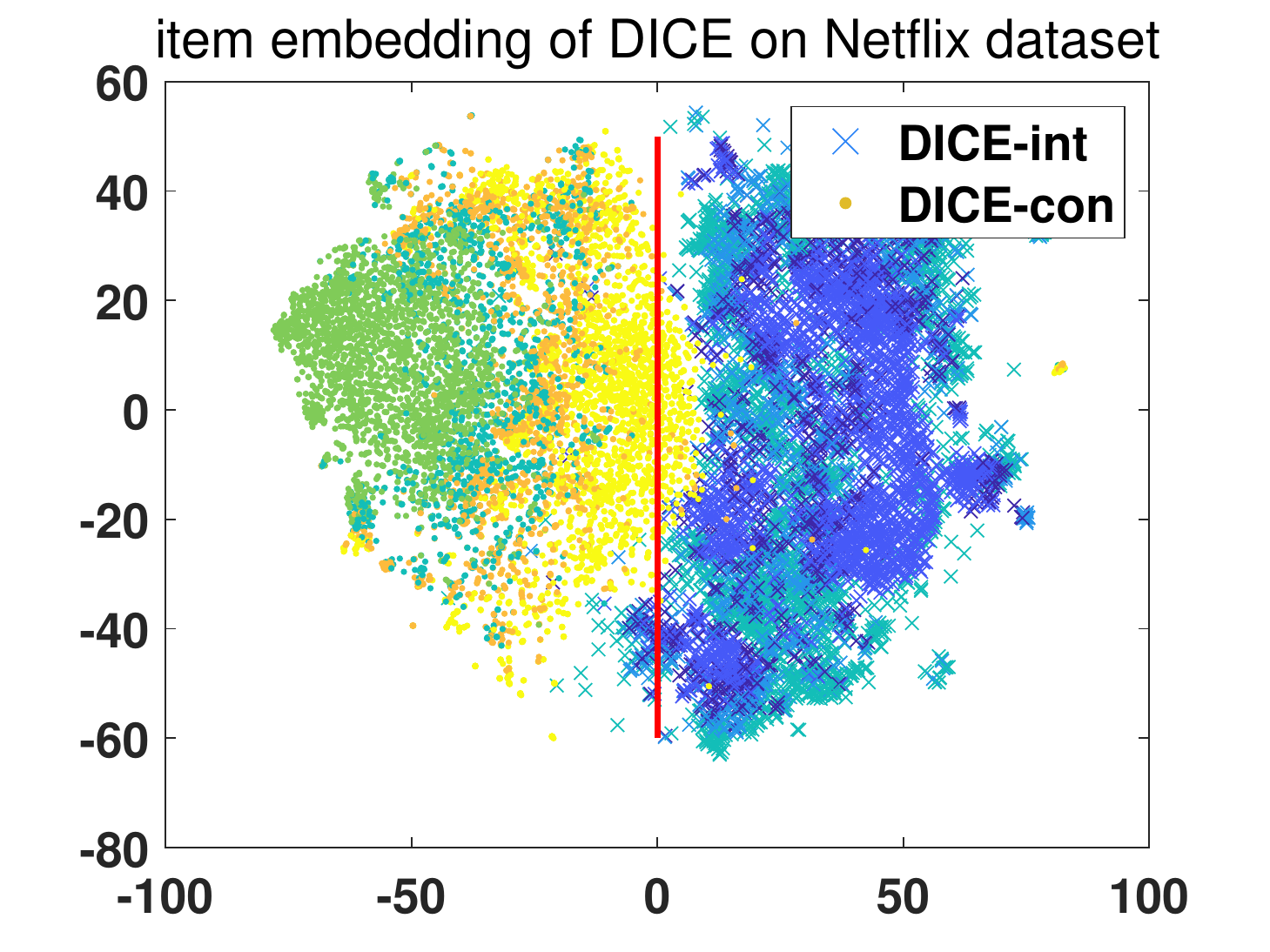}
\end{minipage}
\label{fig::dice_tsne}
\end{minipage}}
\vspace{-10px}
\caption{(a) Overlapped items with ItemPop. Larger IOU means recommendation result is more similar to ItemPop which recommends top popular items. (b) Visualization of the learned item embeddings of DICE on Movielens-10M and Netflix dataset. Interest embeddings are represented by crosses, and conformity embeddings are represented by dots.}
\vspace{-10px}
\end{figure*}

\subsubsection{\textbf{Interpretability based on Disentangled Embedding}} We investigate the quality of embedding disentanglement in DICE. As there is ground-truth for popularity, which serves as a pseudo proxy for conformity. We first study whether conformity embeddings capture the desired cause. Here we introduce another two versions of the framework, DICE-int and DICE-con. They only use interest or conformity embeddings for recommendation, respectively. Note that in DICE we concatenate the two embeddings. We compare the overlapped recommended items of all methods with ItemPop, which recommends the top popular items. Intersection Over Union (IOU) is used as the metric. Figure \ref{fig::overlap} illustrates the results on Movielens-10M dataset. We observe that using conformity embeddings greatly simulates the ItemPop algorithm, and the overlapped items even surpass 50\% when TopK is above 40. Compared with other baselines like IPS and CausE with IOU less than 20\%, DICE-con is much more similar to ItemPop, which validates that conformity embeddings indeed capture the desired cause. The IOU value of DICE-con around 0.5 demonstrates that users tend to confirm with popular items, but different users have their own variance in conformity. If all the users are of the same conformity towards popular items, the IOU value would be close to 1. On the other hand, there is almost no overlapped items between DICE-int and ItemPop, proving that conformity information is almost fully squeezed out from interest embeddings. Therefore, interpretations for interest and conformity can be made based on corresponding embeddings.

Besides calculating the similarity with ItemPop, we visualize the learned item embeddings in DICE using t-SNE \cite{maaten2008visualizing}. Figure \ref{fig::dice_tsne} shows the learned item embeddings on two datasets, where \textit{crosses} represent interest embeddings and \textit{dots} represent conformity embeddings. With special causal learning design and direct supervision on disentanglement, the two sets of embeddings are far from each other, separated by a linear classifier (red line in the figure). Moreover, we divide all the items to three groups based on their popularity, which are popular, normal and unpopular. In Figure \ref{fig::dice_tsne}, items of different groups are painted in different colors. We observe that conformity embeddings are layered according to item popularity, where items of similar popularity are near in the embedding space. Notice that if we use scalar values, items of the three groups will form three segments in a straight line, which is insufficient to capture the diversity of conformity. On the other hand, with respect to interest embeddings, items of different popularity are mixed with each other. Visualizations of the learned item embeddings illustrate the high quality of disentanglement in the proposed framework. Based on disentangled embeddings, reasonable interpretations can be made, which is crucial for recommendation. We also compare the item embedding quality of DICE with baseline methods, including MF, CausE and IPS. Our proposed methodology learns highly interpretable representations for user conformity, and successfully captures the diversity of conformity by using embeddings instead of scalars.

\subsubsection{\textbf{Robustness under Intervention}} 
\begin{figure}[t]
\begin{minipage}[t]{0.49\linewidth}
\centering
\includegraphics[width=\linewidth]{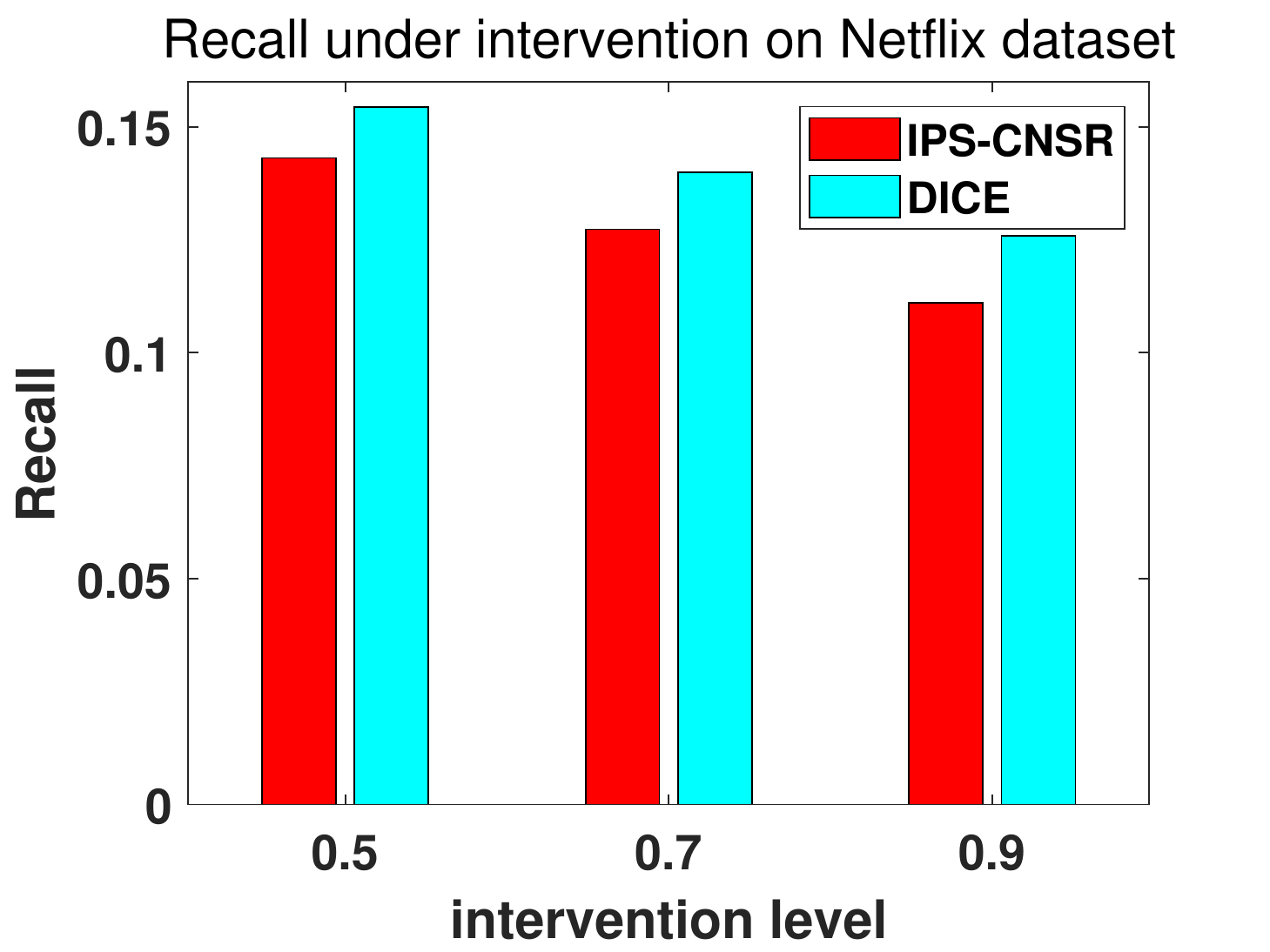}
\end{minipage}
\begin{minipage}[t]{0.49\linewidth}
\centering
\includegraphics[width=\linewidth]{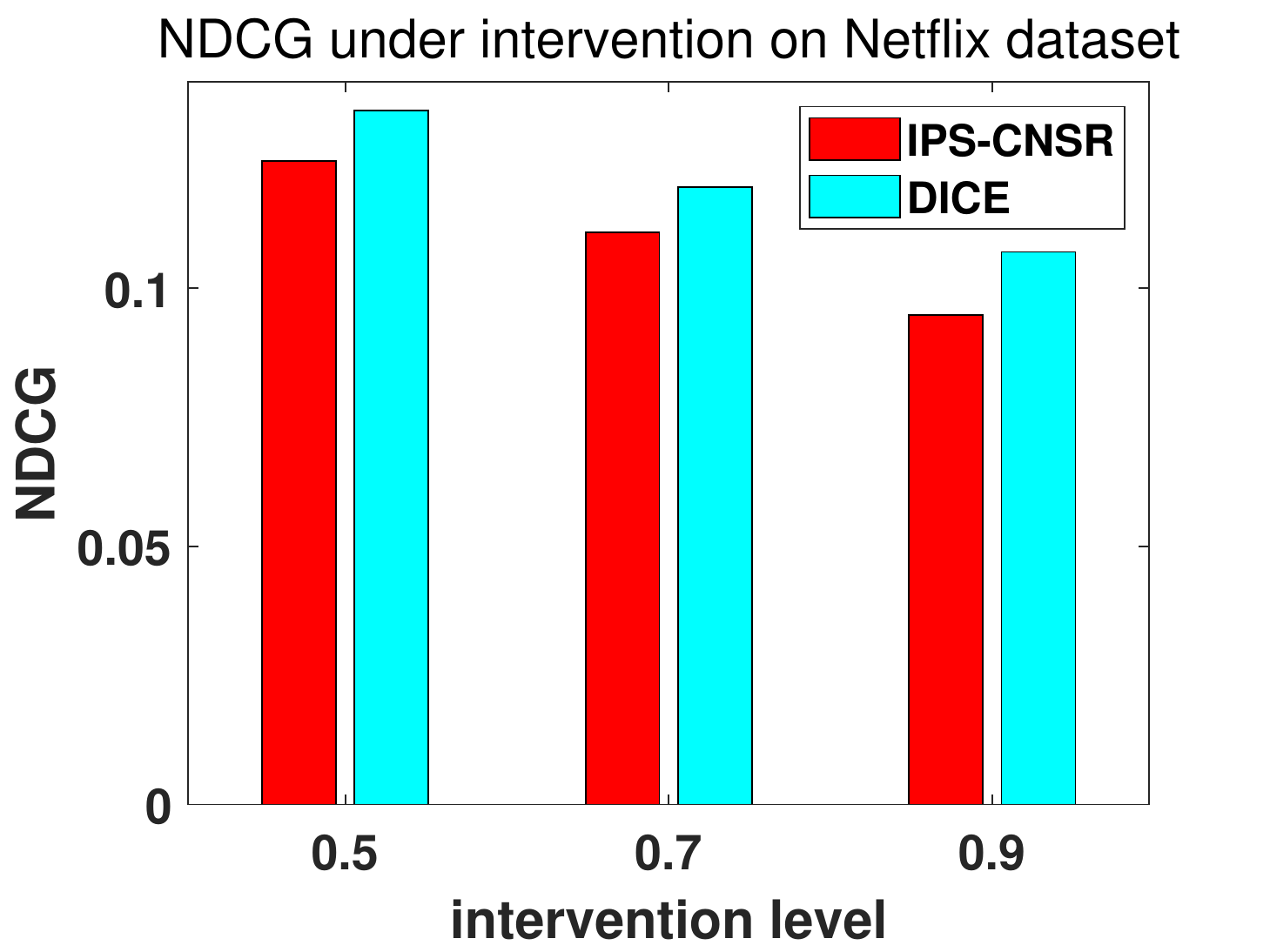}
\end{minipage}
\vspace{-10px}
\caption{Performance comparison between DICE and IPS-CNSR under different levels of intervention.}
\vspace{-10px}
\label{fig::intervention}
\end{figure}

Algorithms that disentangle underlying causes are generally more robust than entangled approaches under intervention \cite{scholkopf2019causality}. In our experiments, we conduct intervention by constructing a different test set which is non-IID with training set, in terms of conformity. Users have access to all items with \textit{equal} probability, rather than seeing more popular items in training set. Specifically, the probability of an instance to be included into test set is the inverse of its corresponding item popularity value. Notice that we cap the probability at 0.9 to avoid too many cold-start items in test set, which controls the strength of intervention. With lower capping value, we impose weaker intervention, hence users are more likely to be exposed with popular items. On the contrary, larger capping value attains stronger intervention and different items receive more equal opportunity to be recommended. Therefore, it provides an elegant way to evaluate the robustness of recommender systems under distinct levels of intervention, by simply changing the capping value. In our experiments, we investigate how the proposed framework performs at different strength of intervention, as well as  state-of-the-art methods. Figure \ref{fig::intervention} shows the results of DICE and IPS-CNSR. We compare the performance of the two approaches with capping value as 0.5, 0.7, and 0.9. The three cases represent quite different interventions on user conformity, since users are more likely to conform towards popular items when capping value is 0.5 due to larger exposure probability for popular items, while they tend to interact according to their real interest when capping value is 0.9 since items are exposed in an almost random manner. Results in Figure \ref{fig::intervention} illustrates that the proposed DICE framework outperforms IPS-CNSR consistently under all degrees of intervention, which proves the robustness of disentangling user interest and conformity.

\subsection{Study on DICE (RQ3)}

Ablation studies on DICE are also conducted to investigate the effectiveness of several components, including negative sampling, conformity modeling, curriculum learning and discrepancy loss.

\begin{table}[t]
\caption{Comparison between the proposed negative sampling strategy PNSM with traditional random strategy on Movielens-10M dataset.}
\centering
{\small
\begin{tabular}{lcccccc}
\toprule
&\multicolumn{3}{c}{Top-K=20}& \multicolumn{3}{c}{Top-K=50} \\
\cmidrule(r{1em}l{1em}){2-4}  \cmidrule(r{1em}l{1em}){5-7}
Name & Recall &  HR & NDCG & Recall &  HR & NDCG \\
\midrule
PNSM & \bf{0.1634} & \bf{0.5197} & \bf{0.1084} & \bf{0.2872} & \bf{0.6975} & \bf{0.1468} \\
RANDOM & 0.1274 & 0.4394 & 0.0843 & 0.2306 & 0.6255 & 0.1160 \\
\bottomrule
\end{tabular}\label{tab::ns}
}
\vspace{-10px}
\end{table}

\subsubsection{\textbf{Impact of Negative Sampling}} As introduced previously, we adopt \textbf{P}opularity based \textbf{N}egative \textbf{S}ampling with \textbf{M}argin (PNSM) to gain high confidence of our causal models. Specifically, when the popularity gap between negative item and positive item is sufficiently large, those derived inequalities on interest and conformity would hold true with high probability. Therefore, we sample items that are significantly more or less popular than the positive item. We require that the popularity gap is larger than a margin value.

In this section, we compare PNSM with the commonly used fully random negative sampling strategy. Table \ref{tab::ns} shows the results on Movielens-10M dataset. We observe that popularity based negative sampling with margin significantly outperforms random negative sampling. Specifically, Recall and NDCG of PNSM are better than RANDOM with over 20\% improvements. PNSM also improves Hit Ratio@20 and Hit Ratio@50 by over 10\%. Results of PNSM and RANDOM verify that sampling negative items with large popularity margin is crucial in the proposed framework. It is reasonable since the proposed causal learning methodology depends on those derived inequalities from causal model \ref{eq::additive}, which will hold true with high probability when the negative items are significantly more or less popular than the positive one.

\subsubsection{\textbf{Impact of Conformity Modeling}} 
\begin{figure}[t]
\begin{minipage}[t]{0.49\linewidth}
\centering
\includegraphics[width=\linewidth]{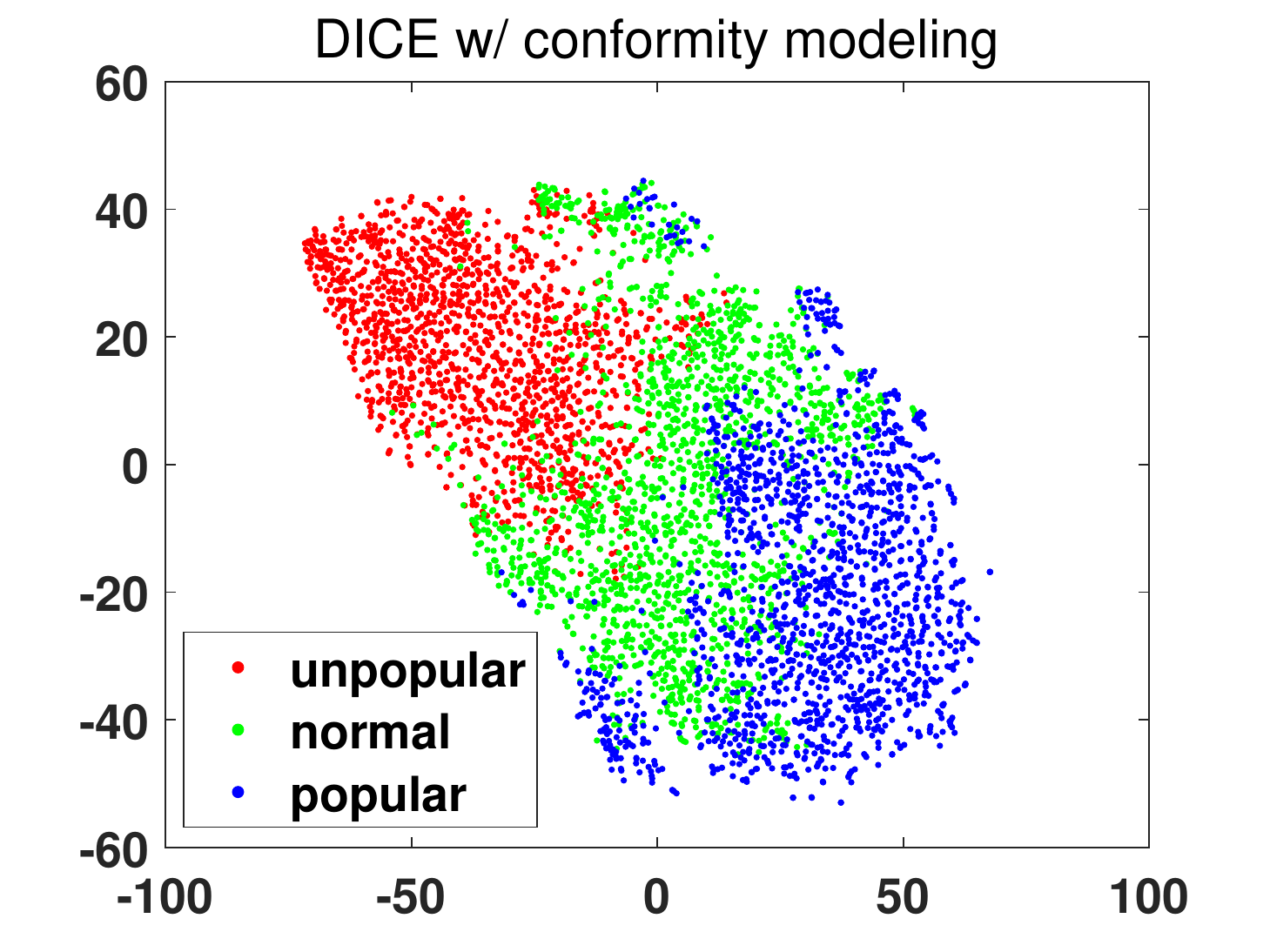}
\end{minipage}
\begin{minipage}[t]{0.49\linewidth}
\centering
\includegraphics[width=\linewidth]{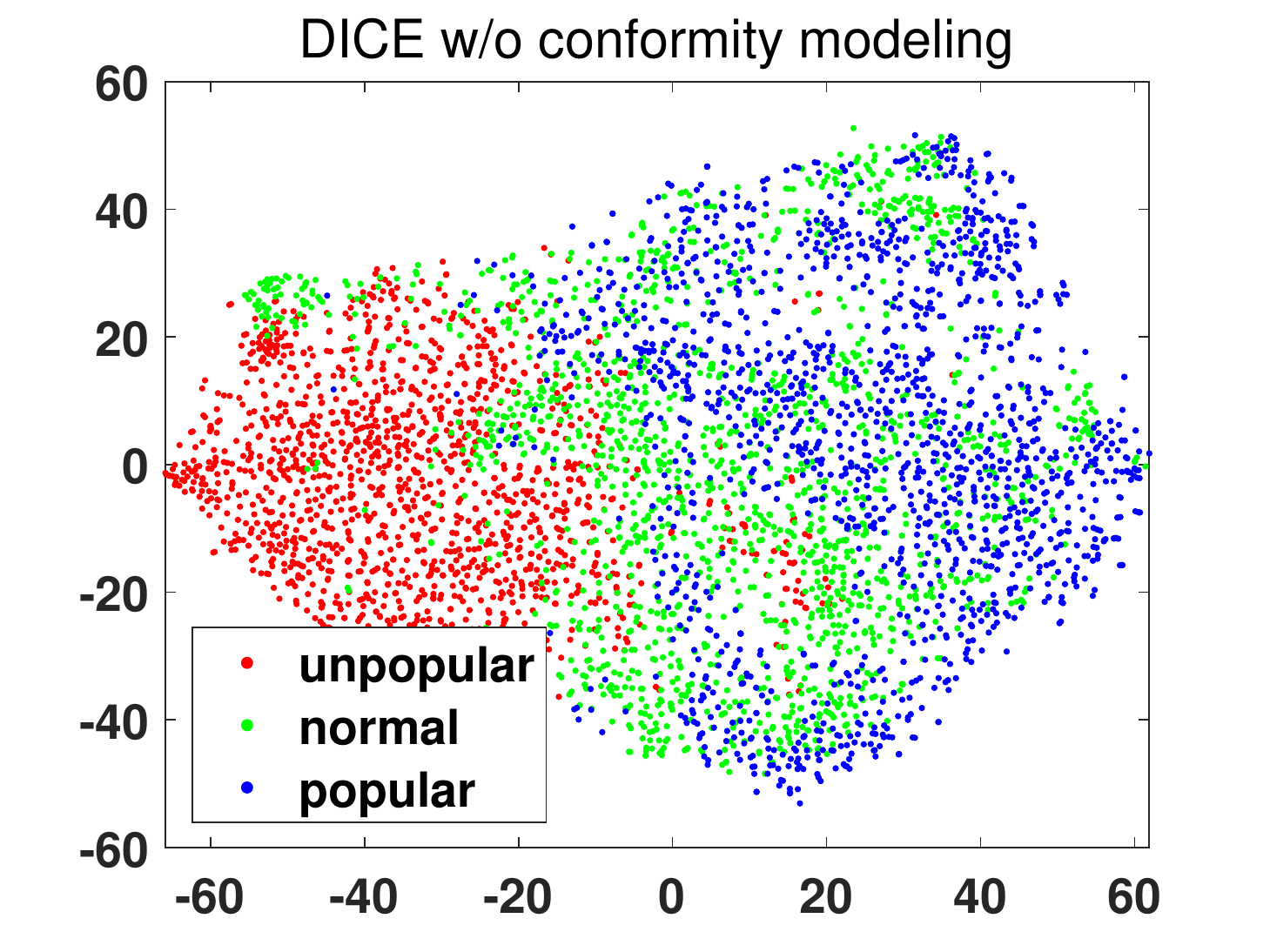}
\end{minipage}
\vspace{-10px}
\caption{Visualization of the learned item embeddings in DICE on Movielens-10M dataset with and without conformity modeling.}
\vspace{-10px}
\label{fig::dice_pop_nopop}
\end{figure}

We also investigate the effect of conformity modeling in DICE. Specifically, we remove the conformity modeling task in DICE and compare it with the full version of DICE. We found that recommendation performance does not decrease much, however, removing conformity modeling task indeed influences the learned embeddings. Figure \ref{fig::dice_pop_nopop} illustrates the learned conformity embeddings in DICE with and without conformity modeling task. We observe that in DICE with conformity modeling task, embeddings are layered according to item popularity, and items of similar popularity are near in the embedding space. However, when we remove the conformity modeling task, the distribution of conformity embeddings becomes messy, and there are more \textit{outliers} in all groups. To be specific, popular items and normal items tend to overlap with each other in the embedding space. Meanwhile, there are also a fraction of normal items lying in the layer of unpopular items.

Conformity modeling task leverages the popularity gap between positive item and negative item to learn pairwise relationships using separate embeddings. From the embedding visualization in Figure \ref{fig::dice_pop_nopop}, we can confirm the effect of conformity modeling task in DICE on learning high-quality interpretable representations.

\begin{figure}[t]
\subfigure[]{
\begin{minipage}[t]{0.47\linewidth}
\centering
\label{fig:mini:subfig:a}
\includegraphics[width=\linewidth]{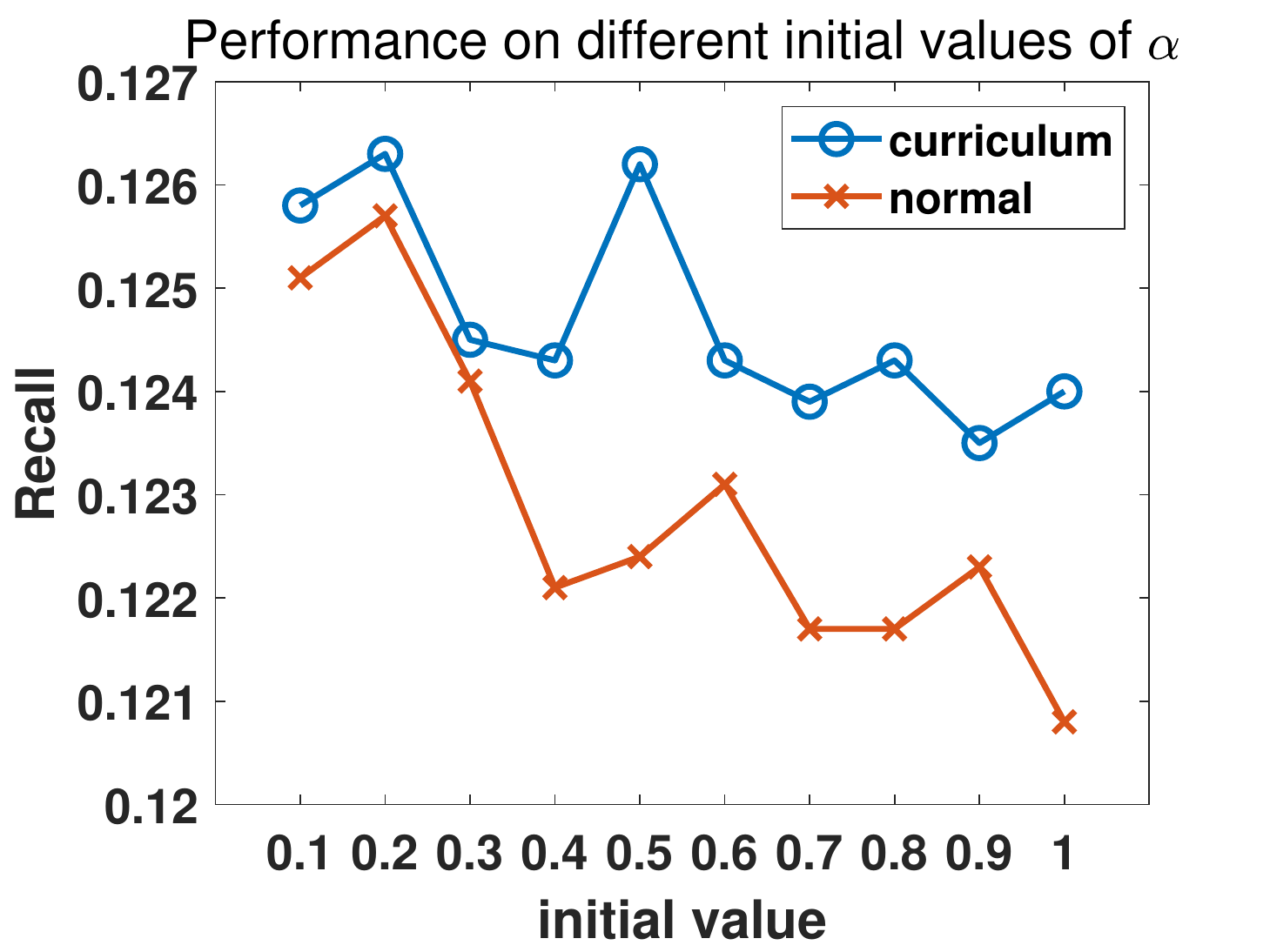}
\label{fig::curriculum}
\end{minipage}}
\subfigure[]{
\begin{minipage}[t]{0.47\linewidth}
\centering
\label{fig:mini:subfig:b}
\includegraphics[width=\linewidth]{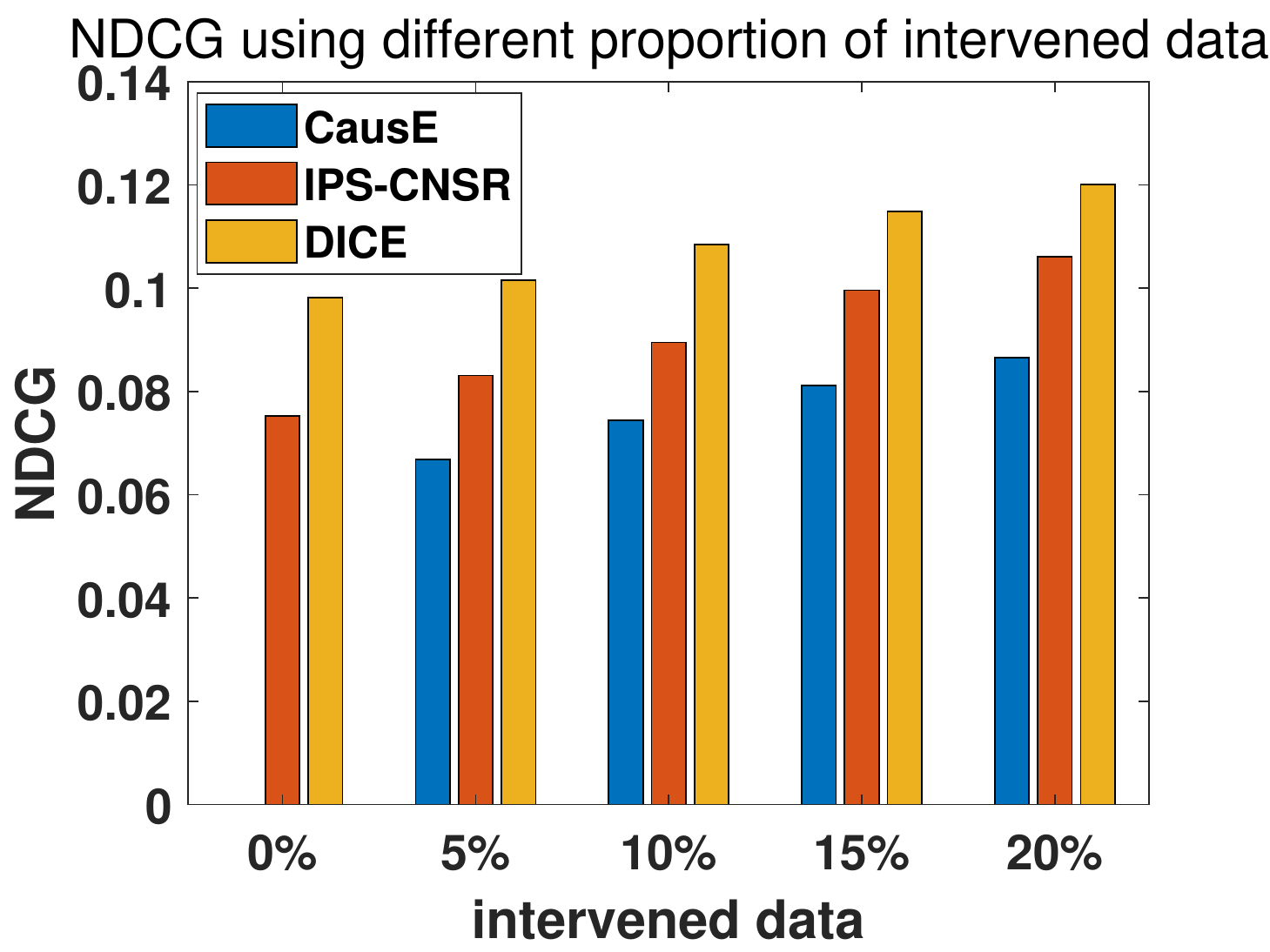}
\label{fig::proportion}
\end{minipage}}
\vspace{-10px}
\caption{(a) Comparison between curriculum learning and normal learning. (b) Performance of different proportion of intervened training data.}
\vspace{-20px}
\end{figure}

\subsubsection{\textbf{Impact of Curriculum Learning}} In the proposed framework, we adopt multi-task curriculum learning to aggregate different causes. Specifically, we make several hyper-parameters adaptive to form a easy-to-hard curriculum for causal learning. These hyper-parameters include loss weight $\alpha$ and negative sampling margin values $m_{up}$ and $m_{down}$. As we train the causal embeddings, we decay these hyper-parameters by a factor of 0.9 to increase the difficulty. In experiments, we initialize these hyper-parameters with different values, and investigate the effect of curriculum learning. Figure \ref{fig::curriculum} shows the results of curriculum learning and normal learning on different initial values of loss weight $\alpha$. We can observe that curriculum learning is consistently better than normal cases. Meanwhile, curriculum learning is not sensitive to initial value due to the easy-to-hard decaying strategy, while normal training without adaptive hyper-parameters is not as stable as curriculum learning and performance drops at large $\alpha$ values.

\subsubsection{\textbf{Impact of Discrepancy Loss}}
We provide three options for discrepancy loss, L1-inv, L2-inv and \textit{dCor}. We examine the three candidates on two datasets with two backbones. Overall, \textit{dCor} attains better performance than L1-inv and L2-inv with over 2\% improvements. However, \textit{dCor} relies on heavy matrix computations which is much more time-consuming than L1-inv and L2-inv. Specifically, training with \textit{dCor} (about 100s per epoch) as discrepancy loss is much slower than L1-inv and L2-inv (about 44s per epoch), which means L1-inv and L2-inv might be more appropriate for large scale applications.

\subsection{Study on Intervened Training Data (RQ4)}

In previous experiments, all the algorithms are trained with a large fraction of normal data (60\%) and a small fraction of intervened data (10\%). Adding extra intervened data is not only a hard requirement of certain baseline method (CausE), but also reduces the difficulty of causal learning. However, intervened data is often too expensive to obtain in real-world recommender systems, \textit{e.g.} random recommendation policy will greatly damage user experience. Therefore, in this section, we investigate how different algorithms perform when we change the proportion of intervened training data, and we also include the most challenging task of not using any intervened data for training. Figure \ref{fig::proportion} demonstrates the performance of DICE, CausE and IPS-CNSR using different proportion of intervened data. Without surprise all the methods got improved performance when we add more intervened data into training set, since it allows models to get access to intervention information which is more similar to test cases. Meanwhile, the proposed DICE framework achieves remarkable improvements against baselines in all cases from 0\% to 20\%. The proposed DICE framework can still disentangle interest and conformity with even no intervened data, and outperforms other baselines significantly. Notice that there is no result for CausE at 0\% since CausE requires intervened training data.

To summarize, we conduct extensive experiments to evaluate the performance of DICE. We compare it with state-of-the-art baseline methods under non-IID circumstances, and DICE outperforms other methods with significant improvements. We emphasize that it is crucial to use embeddings instead of scalars to fully capture the variety of user conformity, which is also proved by experiments against biased MF and biased GCN. Since the main advantages of disentangled algorithms over entangled algorithms are interpretability and robustness, we further conduct experiments to show DICE indeed provides interpretable results and guarantees robustness under intervention. Moreover, we conduct ablation studies to investigate the role of negative sampling, conformity modeling and curriculum learning. At last, we also study the impact of the proportion of intervened training data and different options for discrepancy loss.

\section{Related Work}\label{sec::related}

\textbf{Causal Recommendation}. Existing causal solutions for recommender systems formulate the problem as eliminating popularity bias, from the perspective of items \cite{abdollahpouri2017controlling,bedi2014using,bellogin2017statistical,canamares2017probabilistic,canamares2018should,jannach2015recommenders,marlin2012collaborative,steck2011item}. A bunch of algorithms for unbiased recommendation are proposed in recent literatures, aiming to reduce popularity bias as much as possible \cite{agarwal2019general,bottou2013counterfactual,gilotte2018offline,gruson2019offline,jiang2019degenerate,joachims2017unbiased,liang2016causal,schnabel2016recommendations,wang2018deconfounded,yang2018unbiased}. Among them, Inverse Propensity Scoring (IPS) based methods are mostly adopted and achieve state-of-the-art performance. IPS re-weights each instance as the inverse of corresponding item popularity value, thus popular items are imposed lower weights, while the long-tail items are boosted. IPS guarantees zero bias, however, it is with high variance. A series of variants have been proposed to attain more stable results based on IPS. Bottou \textit{et al.} \cite{bottou2013counterfactual} add max-capping on IPS value, Gruson \textit{et al.} \cite{gruson2019offline} further add normalization, and smoothing and re-normalization are also added to reduce the variance of IPS\cite{gruson2019offline}. IPS and its variants attain unbiased or low-biased recommendation, only from the perspective of items, while ignoring the variety of users' conformity. Imposing different weights is insufficient to comprehensively capture user conformity, since it inherently depends on both user and item.

Besides IPS, Bonner \textit{et al.} \cite{CausE} proposed CausE that performs two MF on a large biased dataset and a small unbiased dataset respectively. L1 or L2 regularization are exploited to force the two factorized embeddings similar with each other. However, conformity is still not taken into consideration in CausE. In recommendation with explicit feedback (\textit{e.g.} rating prediction), Sinha \textit{et al.} \cite{deconvolve} decomposed observed ratings to the union of real ratings and recommender influence. With several strong assumptions, they attained a closed-form solution to recover real ratings from observational ratings based on SVD. However, these assumptions turn out to be invalid in the more prevalent implicit feedback setting. 

Unlike aforementioned approaches that ignore user conformity and bundle different causes into unified representations, our approach achieves causal recommendation with disentangled embeddings for user interest and conformity. To our knowledge, our proposed methodology is the first attempt to tackle the causal recommendation problem from the perspective of users, attaining superior robustness and interpretability by disentangling user interest and conformity.

\noindent\textbf{Disentangled Representation Learning}. Learning representations in which different semantics are disentangled is crucial for robust use of neural models \cite{bengio2019meta,locatello2019challenging,scholkopf2019causality,suter2018robustly}. Existing approaches mainly focus on  computer vision \cite{dupont2018learning,gidaris2018unsupervised,higgins2017beta,hsieh2018learning,kingma2013auto}. For example, $\beta$-VAE \cite{higgins2017beta} learns interpretable representations from raw images in an unsupervised manner. Disentangled representation learning in recommender systems was not explored until recently \cite{ma2019learning,wang2020dgcf,chen2020bias}. Ma \textit{et al.} \cite{ma2019learning} proposed to use Variational Auto-Encoder to disentangle macro-level concept such as intention on different items, and disentangle micro-level factors like color or size of an item. Wang \textit{et al.} \cite{wang2020dgcf} utilized Graph Convolutional Networks to learn disentangled representations for different latent user intentions. These methods decompose user intent into finer granularity, such as the brand or color of an item, while ignoring user conformity, which is essential for recommendation.

\section{Conclusion And Future Work}\label{sec::conclusion}

In this paper, we propose a general framework for disentangling user interest and conformity for recommendation with causal embedding. We develop a concise additive causal model and formulate this model with both causal graph and SCM. Separate embeddings are adopted for interest and conformity according to the proposed SCM. We extract cause-specific data from observational interactions and train different embeddings with different cause-specific data to achieve disentanglement between interest and conformity. The two causes are aggregated and balanced by multi-task curriculum learning. Based on concise and reasonable causal models, DICE consistently outperforms state-of-the-art algorithms with remarkable improvements. Experiments show that DICE is more robust under non-IID circumstances, compared with other baselines. Analysis on disentanglement demonstrates that user interest and conformity are largely independent in the two sets of embeddings. The learned embeddings are of high quality and interpretability, which is promising to explore novel applications using the learned disentangled representations.

DICE decomposes each click interaction into two causes, interest and conformity. A particular meaningful direction for future work is extending DICE to include finer level of causes. For example, the macro-level cause interest could be further divided into micro-level causes such as intentions towards the brand, price or color of items. Overall, we believe disentangling interest and conformity opens new doors for understanding user-item interactions of recommender systems.

\begin{acks}
This work was supported in part by The National Key Research and Development Program of China under grant 2020AAA0106000, the National Natural Science Foundation of China under U1936217,  61971267, 61972223, 61941117, 61861136003, U19A2079.
\end{acks}

\bibliographystyle{ACM-Reference-Format}
\bibliography{bibliography}


\end{document}